\documentclass[twocolumn,amsmath,aps,nofootinbib,superscriptaddress]{revtex4}
\usepackage{graphicx}
\usepackage{subfigure}
\usepackage{bm}
\usepackage{amsmath}
\usepackage{xcolor}
\usepackage{amssymb}
\usepackage{mathrsfs}

\makeatletter
\newcommand{\rmnum}[1]{\romannumeral #1}
\newcommand{\Rmnum}[1]{\expandafter\@slowromancap\romannumeral #1@}
\makeatother

\begin{document}

\title{Evolution of linear wave dark matter perturbations in the radiation-dominant era}

\author{Ui-Han Zhang}
\affiliation{Department of Physics, National Taiwan University, Taipei 10617, Taiwan}

\author{Tzihong Chiueh}
\email{chiuehth@phys.ntu.edu.tw}
\affiliation{Department of Physics, National Taiwan University, Taipei 10617, Taiwan}
\affiliation{National Center for Theoretical Sciences, National Taiwan University, Taipei 10617, Taiwan}


\begin{abstract}
\label{abstract}
Linear perturbations of the wave dark matter, or $\psi$ dark matter ($\psi$DM), of particle mass $\sim 10^{-22}$eV in the radiation-dominant era are analyzed, and the matter power spectrum at the photon-matter equality is obtained.  We identify four phases of evolution for $\psi$DM perturbations, where the dynamics can be vastly different from the counterparts of cold dark matter (CDM).  While in late stages after mass oscillation long-wave $\psi$DM perturbations are almost identical to CDM perturbations, some subtle differences remain, let alone intermediate-to-short waves that bear no resemblance with those of CDM throughout the whole evolutionary history. The dissimilarity is due to quantum mechanical effects which lead to severe mode suppression. We also discuss the axion model with a cosine field potential.  The power spectrum of axion models are generally almost identical to those of $\psi$DM, but in the extreme case when the initial axion angle is near the field potential top, this axion model predict a power excess over a range of wave number and a higher spectral cutoff than $\psi$DM as if $\psi$DM had a higher particle mass.
\end{abstract}

\maketitle
\section{Introduction}
\label{sec: Introduction}

A new form of dark matter, so-called $\psi$DM, wave dark matter or fuzzy dark matter, that consists of extremely light particles of typical particle mass, $10^{-22}$eV \cite{Hu2000, Schive2014, MP2015, CS2016, HuiarXiv2016}, has been noted to exhibit nice, yet peculiar, features for explaining recent observational results. For example, it produces finite density cores for Milky Way dwarf spheroid galaxies to reside in \cite{Schive2014,LM2014}, an observational fact \cite{Moore1994, Goerdt2006, Gilmore2007, WP2011, AAE2013} that has been puzzling cold dark matter (CDM) proponents for some time.  This core in the $\psi$DM scenario results from the uncertainty principle of these extremely light particles, which yields quantum pressure and avoids the central density singularity predicted by CDM \cite{NFW1997}.  Other peculiar features of $\psi$DM arise from the linear matter spectrum described below.

The $\psi$DM spectrum is known to be such that it resembles CDM spectrum in long wavelengths and becomes heavily suppressed in short wavelengths \cite{Hu2000}.  The spectral suppression, due also to the quantum pressure, not only occurs already in the radiation-dominant era but occurs sharply at some transition wavenumber, much sharper than the suppression of the short-wavelength spectrum of warm dark matter (WDM) \cite{Bode2001}. This spectral feature on one hand leads to paucity of Milky Way satellite galaxies \cite{Klypin1999, Moore1999}, which is closely related to the too-big-to-fail problem\cite{Boylan-KolchinBullockKaplinghat2011, Papastergis2015}, and on the other hand postpones first galaxy formation thereby resulting in delay reionization \cite{MarshSilk2014, Schive2016}.  These nice features are not attainable for the WDM model \cite{Maccio2012}. The transition wavenumber, which we call the critical wave number $k_c$, is generally smaller than, but of the same order of, the slowly evolving Jeans wave number $k_J$ of $\psi$DM in the matter-dominant era \cite{Hu2000, WooChiueh2009}. It is therefore of importance to understand how such a spectral transition occurs dynamically and what makes the connection between $k_c$ and $k_J$.

The evolution of the linear $\psi$DM perturbation in the matter-dominant era has been analyzed in \cite{WooChiueh2009}, in which it gave insights as to why the long-wavelength power resembles to the CDM model and how the short-wave power is suppressed due to the presence of quantum Jeans length. While the perturbation dynamics of matter-dominant regime is straightforward, the dynamics in the radiation-dominant regime is much more complex and must be solved numerically \cite{Hlozek2015, Marsh2016}. Unfortunately, the numerical solution offers limited insights to the dynamics. A fluid approach \cite{Marsh2016} for sub-horizon perturbations has also been explored, which can be simple enough to permit detailed analyses and sheds some lights on the perturbation dynamics. However, important dynamics that shapes the final spectrum turns out to occur when perturbations are super-horizon where the fluid approach fails. Hence the fluid approach can only partially answer the questions we intend to address.

Note that $\psi$DM consists of many free bosons condensed in a quantum ground state, a Bose-Einstein condensate, for which they remain phase coherent over an astronomical distance to share the same wave function.  However, there is a problem. Lacking causal contacts, free bosons are not capable of achieving phase coherence.  Nonlinearity is needed to make these bosons interact and become phase-locked. In this regard, the axion mechanism offers a plausible solution to the phase-locking problem, for example \cite{ChiueharXiv2014, HuiarXiv2016}. The axion model can provide strong boson coupling early on and sets free these bosons at a later time; in other words, the phase-lock mechanism takes place in the very early time and since then these bosons remain phase-coherent even after they become free.  Due to this connection to the initial condition of $\psi$DM, it is also important to extend the free-particle model of $\psi$DM and examine the axion model.

The organization of this paper is as follows. Sec. (\ref{sec: Governing Equations}) introduces the relevant equations for later analyses. We also give the fluid formulation from field equations and show that it generally cannot be evolved.  Sec. (\ref{sec: Passive Evolution and Asymptotic Solutions}) contains our main results, where four asymptotic phases are identified, analyzed and approximate solutions obtained.  We make clear the distinction between the passive evolution and the evolution of full treatment in Sec. (\ref{sec: Evolution of Perturbations in Full Treatment}), where solutions of both will be presented and compared.  We identify a special solution common at the boundary of the four asymptotic phases and define a critical wavenumber $k_c$ which marks the spectral transition in Sec. (\ref{sec: Critical mode and Sub-Horizon Dynamics}).  The adiabatic condition for super-horizon modes is examined in Sec. (\ref{sec: Adiabatic Perturbations}) and found to be valid instantaneously for this $\psi$DM problem without the need of time average over the fast mass oscillation.   We examine the axion model and identify its similarities to and differences from the free-particle model in Sec. (\ref{sec: Axion Model}).  Conclusions are made in Sec. (\ref{sec: conclusions}).  In Appendix (\ref{app: passive evolution}), we give full solution of passive evolution.  In contrast to the main text in which we obtain solutions to the approximate equations in various asymptotic phases, we derive in Appendix (\ref{app: passive evolution}) approximate solutions to the full solution for the four asymptotic phases.  In Appendix (\ref{app: full treatment evolution}), we turn to addressing equations and analyses for the full treatment of perturbations, including neutrino, baryon and photon perturbations.

Throughout the paper, we take the fiducial particle mass to be $10^{-22}$eV and adopt the Newtonian gauge for the perturbation.  We also let the speed of light, $c$, and the Planck constant $\hbar$ be equal to $1$.

\section{Governing Equations}
\label{sec: Governing Equations}

Standard derivation of the governing equations is given in Appendix (\ref{app: passive evolution}).  Here, we provide notations and essential equations to be used in this paper.

Let $a_*$ and $H_*$ be the scaling factor and Hubble parameter at some given epoch in the radiation era, and define $d\tau= dt/a$ and the Hubble parameter $H\equiv d\ln a/d\tau$.  Since $H\propto a^{-1}$, we have $Ha=H_*a_*$, a fixed constant, and it follows $\tau=(H_*a_*)^{-1} a = H^{-1}$.

Moreover, we decompose the $\psi$DM wave function into a time-dependent background $\Psi(\tau)$ and a space-time dependent perturbation $\psi$ as usual.  The metric perturbation in the Newtonian gauge is denoted as $\phi$.  The quantities $\delta_\gamma$ and $\theta_\gamma$ are the dimensionless energy density perturbation and the velocity potential of the photon fluid.  These perturbed quantities in the comoving coordinate are Fourier transformed in space into plane-wave eigenmodes with the comoving wavenumber $k$ as eigenvalues.

Denote the prime to be $d/d\tau$. The zeroth-order field $\Psi$ obeys
\begin{equation}
\label{equ: zeroth-order dark matter}
\Psi^{''} + 2H\Psi^{'} + m^2 a^2\Psi = 0,
\end{equation}
and the perturbed field $\psi$ obeys
\begin{equation}
\label{equ: first-order dark matter}
\psi^{''}+2H\psi^{'}+(k^2 + m^2a^2)\psi = 4\Psi^{'}\phi^{'}-2m^{2}a^{2}\Psi\phi.
\end{equation}
The radiation perturbation equations are
\begin{equation}
\label{equ: first-order photon fluid 1}
\delta_\gamma^{'}-{4\over 3}(k^2\theta_\gamma+3\phi^{'})=0,
\end{equation}
and
\begin{equation}
\label{equ: first-order photon fluid 2}
\theta_\gamma^{'}=-{\delta_\gamma\over 4}-\phi,
\end{equation}
where $\delta_\gamma\equiv \delta\epsilon_\gamma/\epsilon_\gamma$ with $\epsilon_\gamma$ being the energy density of the radiation fluid, and $\theta_\gamma$ is the perturbed velocity potential of the radiation fluid.  Neutrinos can be regarded as a part of the radiation fluid for super-horizon perturbations.  But upon entering horizon, neutrino perturbations die out sharply as a result of their collisionless nature where collisionless damping prevails. In Sec. (\ref{sec: Evolution of Perturbations in Full Treatment}) we have more discussions on this issue.

The equations for the metric perturbation $\phi$ read
\begin{equation}
\label{equ: metric perturbation equation 1}
\left.\begin{aligned}
& -k^2\phi-3H(\phi^{'}+H\phi)= \\
& 4\pi G \{ [\Psi^{'}\psi^{'}+m^2a^2\Psi\psi-(\Psi^{'})^2\phi]+a^2\epsilon_\gamma\delta_\gamma \},
\end{aligned}
\right.
\end{equation}
and
\begin{equation}
\label{equ: metric perturbation equation 2}
\phi^{'}+H\phi=4\pi G[\Psi^{'}\psi-a^2(\epsilon_\gamma +P_\gamma)\theta_\gamma].
\end{equation}

One can identify the perturbed energy density as
\begin{equation}
\label{equ: energy perturbation}
a^2\epsilon_\psi\delta_\psi=\Psi^{'}\psi^{'}+m^2a^2\Psi\psi-(\Psi^{'})^2\phi
\end{equation}
on the right-hand side of Eq. (\ref{equ: metric perturbation equation 1}), where
\begin{equation}
\label{equ: energy density}
\epsilon_\psi \equiv [(\Psi')^2+m^2a^2\Psi^2]/(2a^2).
\end{equation}

One can further substitute Eq. (\ref{equ: metric perturbation equation 2}) into Eq. (\ref{equ: metric perturbation equation 1}) to obtain a simplified equation. Recognizing the gauge covariant energy perturbation
\begin{equation}
\label{equ: covariant energy}
\epsilon_\alpha\Delta_\alpha = \epsilon_\alpha\delta_\alpha-3H(P_\alpha+\epsilon_\alpha)\theta_\alpha, \text{  } \alpha=\gamma,\text{ } \psi,
\end{equation}
and the momentum potential
\begin{equation}
\label{equ: momentum perturbation}
\theta_\psi = -{\Psi^{'}\psi\over a^2(P_\psi+\epsilon_\psi)},
\end{equation}
with
\begin{equation}
\label{equ: pressure}
P_\psi\equiv[(\Psi^{'})^2- m^2a^2\Psi^2]/(2a^2),
\end{equation}
it follows that Eq. (\ref{equ: metric perturbation equation 1}) becomes
\begin{equation}
\label{equ: Poisson equation}
-k^2\phi=4\pi G a^2(\epsilon_\psi \Delta_\psi + \epsilon_\gamma\Delta_\gamma).
\end{equation}
The Poisson equation with gauge covariant sources is recovered.

Although the fluid description of $\psi$DM is not useful in general, it is illuminating to find out its difference from that of CDM. We multiply Eq. (\ref{equ: first-order dark matter}) by $a\Psi^{'}$ and add Eq. (\ref{equ: zeroth-order dark matter}) multiplied by $a\psi^{'}$.  As the perturbed pressure as
\begin{equation}
\label{equ: pressure perturbation}
\delta P_\psi = a^{-2}\Psi^{'}\psi^{'}- m^2\Psi\psi - a^{-2}(\Psi^{'})^2\phi,
\end{equation}
we find the perturbed energy equation
\begin{equation}
\label{equ: energy density equation}
\delta_\psi^{'}+ [ln(a^3 \epsilon_\psi)]^{'}\delta_\psi+3H{\delta P_\psi\over\epsilon_\psi} -(1+{P_\psi\over\epsilon_\psi})(k^2\theta_\psi+3\phi^{'})=0.
\end{equation}
For the perturbed momentum equation, we multiply Eq. (\ref{equ: zeroth-order dark matter}) by $\psi$ and obtain
\begin{equation}
\label{equ: momentum density equation}
\theta_\psi^{'} + \{ln[(P_\psi + \epsilon_\psi) a^4] \}^{'}\theta_\psi = - {\delta P_\psi\over P_\psi+\epsilon_\psi}-\phi.
\end{equation}

Equations (\ref{equ: energy density equation}) and (\ref{equ: momentum density equation}) are actually quite general and valid for any standard field potential.  They become the radiation fluid perturbations when $(\delta P, P) = (1/3)(\epsilon \delta, \epsilon)$ with $\epsilon \propto a^{-4}$ and also the CDM perturbations when $\delta P = P =0$ with $\epsilon\propto a^{-3}$, where the "$\psi$" index is replaced by the photon index and the CDM index, respectively.  Note that Eqs. (\ref{equ: energy density equation}) and (\ref{equ: momentum density equation}) need equations of state to relate $P$ to $\epsilon$ and $\delta P$ to $\epsilon \delta$. These relations are readily available for standard fluids, but generally far from trivial here.  In fact, one must solve the zero-order and perturbed field equations, Eqs. (\ref{equ: zeroth-order dark matter}) and (\ref{equ: first-order dark matter}), to construct $P_\psi$ and $\delta P_\psi$ expressed in terms of $\epsilon_\psi$ and $\delta\epsilon_\psi$.  Therefore without solving for the field, the fluid description of the field is generally of no practical use.

\section{Passive Evolution and Asymptotic Solutions}
\label{sec: Passive Evolution and Asymptotic Solutions}

Passive evolution of dark matter can be a good approximation of the dark matter dynamics in the radiation-dominant era, where the metric perturbation is governed entirely by the radiation fluid, which includes neutrinos.  Neutrinos are actually collisionless particles, and their perturbations behave identical to radiation perturbations before entering horizon but die out rapidly once they enter horizon \cite{MaBertschinger1995}.  We approximate neutrinos always as a part of the radiation fluid for passive evolution, and will show in the next section that this crude treatment is still a good approximation for dark matter perturbations of main interest.

The right-hand side of Eq. (\ref{equ: first-order dark matter}) can thus be approximated as external sources independent of $\psi$ for passive evolution.  This assumption greatly simplifies the following asymptotic analyses, bringing out the details of the differences between the CDM perturbation and the $\psi$DM perturbation caused by the introduction of a finite particle mass $m$ in Eq. (\ref{equ: first-order dark matter}). In the limit of $m\to \infty$, the $\psi$DM perturbation should recover the CDM result.  With this approach, analytical expressions for source terms in Eq. (\ref{equ: first-order dark matter}) can actually be obtained and therefore the solution $\psi$ can be integrated by the Green's function method.  The full solution however involves confluent hypergeometric functions and is not illuminating to be given in the main text, and hence we present it in Appendix (\ref{app: passive evolution}).

\begin{figure}
\includegraphics[scale=0.35, angle=270]{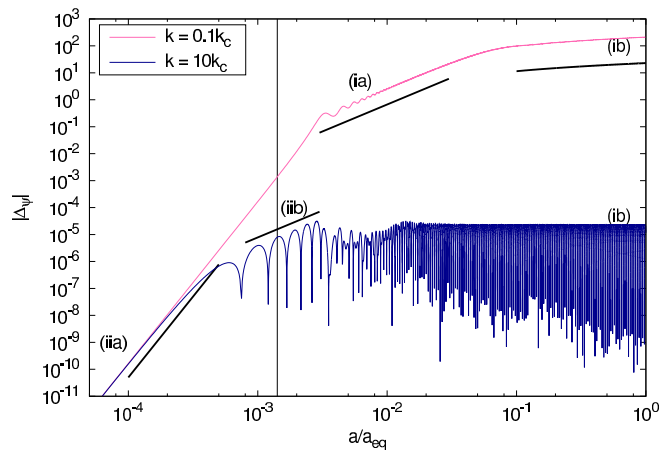}
\caption{Two cases of gauge covariant $\psi$DM energy density perturbations $\Delta_{\psi}$ with passive evolution for the very long and very short wave modes. The horizontal axis is $a/a_{eq}$ where $a_{eq}$ is the scale factor at radiation-matter equality, and the particle mass is chosen to be $10^{-22}$eV. There are four asymptotic phases as labeled.  The vertical line marks the beginning of mass oscillation. The bold lines have log-slopes $6$, $2$, $\sim 0$, respectively from left to right.}
\label{fig: compare_passive_active_free_evolution}
\end{figure}

Plotted in Fig. (\ref{fig: compare_passive_active_free_evolution}) is the passive time evolutions of two distinct modes with mode numbers $k = 0.1k_c$ and $k = 10k_c$, where $k_c$ is the wave number of the critical mode that enters horizon at the onset of mass oscillation.  (More discussions on $k_c$ are given in Sec. (\ref{sec: Critical mode and Sub-Horizon Dynamics}).)  There exist four asymptotic phases of evolution as shown in Fig. (\ref{fig: compare_passive_active_free_evolution}), and this plot sets the stage for discussions to follow.

The evolution of $\psi$DM perturbations has a natural dividing line, the onset of mass oscillation. Prior to the mass oscillation, $\psi$DM is like an inflaton and after that, $\psi$DM is like CDM influenced to various degrees by the additional quantum pressure.  Since the latter is closely resembled the familiar CDM, we shall discuss the second phase first. Little surprise arises in the second phase, given that we know the perturbed $\psi$DM dynamics in the matter-dominant era \cite{WooChiueh2009}.  The new features of $\psi$DM is actually given by the earlier phase prior to mass oscillation, resulting in a sharp transition of the $\psi$DM spectrum.

\bigskip

(\rmnum{1}) After mass oscillation $2H, k \ll ma$

\medskip

For long wave ($k \ll ma$) perturbations well after the onset of mass oscillation, $ma \gg 2H$, the dynamical system is like a driven damped oscillator near resonance, with the main frequency $m$ and detuning frequency $k$. Equation (\ref{equ: first-order dark matter}) can be simplified by extracting the common mass oscillation factor in all terms and is approximated by a complex amplitude equation, for which the detuning frequency $k$ appears in the equation. The simplification procedure goes as follows.  Let $\psi\equiv A\Psi_r+B\Psi_s$, where $\Psi_r$, the regular solution of $\Psi$, and $\Psi_s$, the singular solution that diverges at $\tau \to 0$, are given in Appendix (\ref{app: passive evolution}), $A$ and $B$ are real fields with slow time variations, and the zero-order field $\Psi=\Psi_r$. To the accuracy of $O(2H/ma)$, $(\Psi_r, \Psi_r'/ma)\sim (2\epsilon_\psi/m^2)^{1/2}(\cos(mt-q+\alpha)+O(H^2/m^2a^2), -\sin(mt+7q+\alpha)+O(H^2/m^2a^2))$, with $\alpha$ being a constant phase and $q = 3H/16 ma\propto 1/t$, and $(\Psi_s, \Psi_s'/ma)\sim (2\epsilon_\psi/m^2)^{1/2}(\sin(mt-q+\alpha)+O(H^2/m^2a^2),\cos(mt+7q+\alpha)+O(H^2/m^2a^2))$. One may further define a complex $\hat\psi$, where $\hat\psi = A+iB$, and therefore $\psi = Re[\hat\psi(\Psi_r - i\Psi_s)]\to (2\epsilon_\psi/m^2)^{1/2} \Re[\hat\psi \exp(-i(mt+\alpha))]$ in the limit $H/ma\to 0$.

The amplitude $\hat\psi$ equation can be straightforwardly derived from Eq. (\ref{equ: first-order dark matter}),
\begin{equation}
\label{equ: amplitude equation}
-i\hat{\psi}^{'} +{k^2\over 2ma}\hat{\psi}= -ma\phi.
\end{equation}
This equation is the familiar linearized Schr\"{o}dinger equation for the perturbed field.

Note that in order for Eq. (\ref{equ: amplitude equation}) to be correctly derived, it is essential to keep the small $q$ in $\Psi_r$ and $\Psi_s$.  Otherwise, there would be additional incorrect terms of order $O(H^2 \hat\psi)$ appearing in Eq. (\ref{equ: amplitude equation}), which would have otherwise given an erroneous solution for Phase (\rmnum{1}a).

Eq. (\ref{equ: amplitude equation}) is driven by a source proportional to $\phi$, and the solution consists of a particular integral and a homogeneous solution.  The adiabatic perturbation, which is of interest to us, corresponds to the particular integral of the full equation, Eq. (\ref{equ: first-order dark matter}).  For the asymptotic regimes to be discussed below, certain limits have been taken to simplify the full equation, but this simplification comes with a cost, in decomposing the particular integral of the full equation into the particular integral and the homogeneous solution of the simplified equation.  The homogeneous solution can be regarded as the initial condition of the solution to the simplified equation and assumes a undetermined amplitude, which needs to be fixed by some means, for example, matching to the solution in the previous phase.

\bigskip

(\rmnum{1}a) $k \ll 2H \ll ma$

\medskip

In this regime, the term proportional to $k^2$ in Eq. (\ref{equ: amplitude equation}) can be ignored to the leading order.  Now, as the super-horizon metrics perturbation $\phi=\phi_0[1-(1/30)(k/H)^2+O(k^4/H^4)]$ where $\phi_0$ is a constant (shown in Appendix (\ref{app: passive evolution})), the driving source in the real part of Eq. (\ref{equ: amplitude equation}) grows as $a$.  Thus the particular integral solution $\Im[\hat{\psi}]_p=-(m a/2H)\phi_0\propto a^2$, and $\Re[\hat\psi]_p =0$.  On the other hand the homogeneous solution, $\hat\psi_h$, of Eq. (\ref{equ: amplitude equation}) is a complex constant.

To the leading order, we have
\begin{equation}
\label{equ: deltapsi}
\delta_\psi = 2\Re[\hat\psi] +{{3H}\over{ma}}[2\sin^2(mt+\alpha)-1]\Im[\hat{\psi}]
\end{equation}
(c.f. Eq. (\ref{equ: energy perturbation})), where $\Im[\hat\psi]^{'}=-ma\phi_0$ has been used.  The quantity $\Re[\hat\psi]$ is an arbitrary homogeneous solution $\Re[\hat\psi]_h$, and $\Im[\hat{\psi}]$ is a combination of the particular integral and a constant homogeneous solution.  Finally, the covariant energy perturbation
\begin{equation}
\label{equ: Delta1}
\left.\begin{aligned}
\Delta_\psi & = \delta_\psi - {{6H}\over{ma}}\Im[\hat\psi]\sin^2(mt+\alpha) \\
            & = 2\Re[\hat\psi] - {{3H}\over{ma}}\Im[\hat\psi].
\end{aligned}
\right.
\end{equation}
Note that while the gauge-dependent $\delta_\psi$ oscillates, the gauge-covariant $\Delta_\psi$ manifestly does not.

It requires some guidance to fix the homogeneous solution $\hat\psi_h$ in Eq. (\ref{equ: Delta1}). Conventionally, matching of solutions in different asymptotic regimes can provide such guidance.  However we adopt a different approach here.  For super-horizon adiabatic perturbation, $\delta_\psi$ obeys the relation:
\begin{equation}
\label{equ: adiabatic condition}
{\langle \delta_i\rangle \over \langle 1+w_i\rangle} = {\langle \delta_j\rangle\over \langle 1+ w_j\rangle}
\end{equation}
for any species $i$ and $j$, where we denote $<...>$ to represent a short-time average to filter out the fast mass oscillation.  (However, see Sec. (\ref{sec: Adiabatic Perturbations}).)  It follows $\langle \delta_\psi\rangle = (3/4)\delta_\gamma$.  From Appendix (\ref{app: passive evolution}), we also know that $\delta_\gamma \sim -2\phi_0$ and hence $\langle\delta_\psi\rangle\sim -(3/2)\phi_0$.  Therefore the constant $\Re[\hat\psi]_h=-(3/4)\phi_0$.  Given $\Im[\hat\psi]_p=-(ma/2H)\phi_0$, we find
\begin{equation}
\label{equ: Delta2}
\Delta_\psi = -{{3H}\over{ma}}\Im[\hat\psi]_h.
\end{equation}
The $\phi$ dependence in $\Delta_\psi$ cancels and only the arbitrary homogeneous solution $\Im[\hat\psi]_h$ survives.

If $\Im[\hat\psi]_h \sim O(\Re[\hat\psi]_h)$, the remaining term of $\Delta_\psi$ is a small quantity of order $2H/ma$ and decays as $a^{-2}$, thus negligible. From the definition of $\hat\psi$, the imaginary part $\Im[\hat\psi]$ is to be multiplied by $\Psi_s$ and the real part $\Re[\hat\psi]$ by $\Psi_r$ to yield the original field $\psi$. The ratio $\Psi_s/\Psi_r$ diverges in early epoches when $2H/ma \to \infty$, and the ratio $\Im[\hat\psi]/\Re[\hat\psi]$ in early time must approach zero.  Therefore at the onset of mass oscillation $2H=ma$ immediately before the perturbation enters the present phase, which is the main contribution of the homogeneous solution, the quantity $\Im[\hat\psi]_h$ is at most of the same order of $\Re[\hat\psi]_h$.   In Appendix (\ref{app: passive evolution}), we show that $\Im[\hat\psi]_h$ is in fact of higher-order smallness compared to $\Re[\hat\psi]_h$.  Therefore the gauge-covariant $\Delta_\psi$ cancels itself to the leading order ($O(1)\phi_0$), and we must consider the next-order contributions to $\Delta_\psi$.

In this phase, there are two types of high-order terms, $O(2H/ma)$ and $O((k/H)^2)$.  The former decays as $a^{-2}$ and the latter grows as $a^2$.  We shall consider the latter. The next order contributions to $\Delta_\psi$ arise from all other terms in Eq. (\ref{equ: first-order dark matter}) but are neglected in Eq.(\ref{equ: amplitude equation}).

A straightforward but lengthy calculation by keeping all terms of order $O(k^2/H^2)$ in Eq. (\ref{equ: first-order dark matter}) reveals that $\Re[\hat\psi]_p = -(7/40)(k/H)^2\phi_0$ and $\Im[\hat\psi]_p = - (ma/2H)\phi_0(1-(1/60)(k/H)^2)$.  The homogeneous solution $\Re[\hat\psi]_h = -(3/4)\phi_0(1 + O((k/H)^2(H/ma)))$ and remains the same as before to the order in question; so is $\Im[\hat\psi]_h$. Substituting the above findings to the second equality of Eq. (\ref{equ: Delta1}), we now have $\Delta_\psi = -(3/8)(k/H)^2\phi_0 = (9/16)\Delta_\gamma \propto a^2$, where the Poisson's equation (Eq. (\ref{equ: Poisson equation})) has been used to bring out $\Delta_\gamma$.

The $\psi$DM gauge-covariant energy perturbation in this phase is identical to the CDM counterpart, for which the growth is independent of the particle mass $m$.  Note that the perturbed field goes through this phase only for sufficiently low-$k$ modes.   Modes with sufficiently high-$k$ skip this phase and directly enter Phase (\rmnum{1}b) to be discussed below.

\bigskip

(\rmnum{1}b) $2H \ll k \ll ma$

\medskip

In this regime, the pressure perturbation can be of dynamical importance despite $\langle P_\psi\rangle = 0$.  Here, $\langle\delta P_\psi\rangle=(\epsilon_\psi/ma)\Re[i\hat{\psi}^{'}-ma\phi]$ from Eq. (\ref{equ: pressure perturbation}).  Compared with the real part of Eq. (\ref{equ: amplitude equation}), one readily recognizes that
\begin{equation}
\label{equ: average pressure pertubation}
\langle\delta P_\psi\rangle = {{k^2}\over {2m^2 a^2}}\epsilon_\psi (\Re[\hat{\psi}])\approx {{k^2}\over {4m^2 a^2}}\epsilon_\psi\langle\delta_\psi\rangle.
\end{equation}
The second equality of Eq. (\ref{equ: average pressure pertubation}) holds because the term proportional to the metric perturbation $\phi$ in $\delta_\psi$ is a high-order term and can be ignored.  Now that with $\langle\delta P_\psi\rangle$ available, the dynamics of perturbation in the sub-horizon regime can be described by fluid equations.

As $\epsilon_\psi \langle\delta_\psi\rangle \gg \langle\delta P_\psi\rangle$, we seem to recover CDM.  But it is not so even when $k^2 \ll m^2 a^2$, since this inequality is only an indication of $\psi$DM becoming non-relativistic.  Non-relativistic dark matter can nevertheless have sufficiently large pressure to counter the gravity for short waves.  Hence only in the long wave limit of this regime $k^2 \ll (m^2 a^2)(2H/ma)$ can $\psi$DM perturbations resemble CDM perturbations, as we shall see below.

In discussions to follow, we shall stick to the field equation for consistency, despite that the fluid equations are also well-defined.  In this regime the metric perturbation $\phi \sim \phi_0 (H/k)^2 \cos(k(\tau-\tau_k)/\sqrt{3})$ as discussed in Appendix (\ref{app: passive evolution}), where $\tau_k$ is the instant for mode $k$ to enter horizon.  Therefore the driving source ($\propto \phi$) of Eq. (\ref{equ: amplitude equation}) is of order $O(a^{-1})$.  It is straightforward to find that the particular integrals $\Re[\hat\psi]_p\sim \phi_0(3/2)(H/k)^2\cos(k(\tau-\tau_k)/\sqrt{3}))$ and $\Im[\hat\psi]_p\sim  -\phi_0(\sqrt{3}maH^2/k^3)\sin(k(\tau-\tau_k)/\sqrt{3})$, which decay as $a^{-2}$ and $a^{-1}$, respectively, and can be ignored.

On the other hand, the homogeneous solution derived below exhibits constant-amplitude oscillations and will dominate the particular integral. It is convenient to choose the reference epoch $a_*=a_m$ and $H_*=H_m$, where $a_m$ and $H_m$ refer to the expansion factor and Hubble parameter at $2H=ma$, or $H_m\equiv ma_m/2$.  Multiplying Eq. (\ref{equ: amplitude equation}) by $a$, we can change the time variable to a dimensionless $\eta = \ln (a/a_m)\equiv\ln(H_m\tau)$.  Equation (\ref{equ: amplitude equation}) can be cast into
\begin{equation}
\label{equ: amplitude equation 1}
-i{d\hat{\psi}_h\over d\eta}+{k^2\over{4H_m^2}}\hat{\psi}_h= 0.
\end{equation}
This equation admits the homogeneous solution $\hat{\psi}_h = g\exp[i(k^2/4H_m^2)(\eta-\eta_k)]$ with $g$ being a complex constant and $\eta_k = \ln(a_k/a_m)$ denoting the duration between the onset of mass oscillation and when mode $k$ enters horizon.  The homogeneous solution is oscillating with a constant amplitude and dominates the particular integral for $a >> a_k$. Thus the perturbed field $\hat{\psi}$ decouples from the gravity in this regime and becomes a free matter wave with a constant frequency $k^2/4H_m^2$ in the $\eta$ space.

In the long-wave limit, where $k^2/4H_m^2 \ll 1$, we can estimate the complex constant $g$ by matching the solution to Phase (\rmnum{1}a), which has a constant $\Re[\hat\psi]\sim O(1)\phi_0$ and a growing $\Im[\hat\psi]\sim O(ma/H)\phi_0 (>>\Re[\hat\psi])$.  We therefore expect that in the limit of small $\eta-\eta_k$ in the present phase, $\Re[\hat\psi] \ll \Im[\hat\psi]$.  This is only possible when $\Re[\hat\psi]\propto \sin[(k^2/4H_m^2)(\eta-\eta_k)]$ and $\Im[\hat\psi]\propto \cos[(k^2/4H_m^2)(\eta-\eta_k)]$.  That is, $g$ is pure imaginary.   Moreover, we can also estimate that $|g|\sim O(H_m^2/k^2)\phi_0$, obtained from the small argument expansion of $\sin[(k^2/4H_m^2)(\eta-\eta_k)]$ for $\Re[\hat\psi]$.   This estimate of $|g|$ is also consistent with $\Im[\hat\psi]$ at the end of Phase (\rmnum{1}a) when $k\to H$, since the factor $ma/H = maH/H^2 \to H_m^2/2k^2$.

To fix the value of $|g|$, we use the CDM limit in Appendix (\ref{app: passive evolution}) with $H/ma \to 0$.  The CDM covariant energy perturbation is $\Delta_{CDM} \to  (c_1 -9 (\eta-\eta_k))\phi_0$, when $c_1$ is a constant of order unity.  On the other hand, the expression of the covariant energy perturbation $\Delta_\psi$ is the same as Eq. (\ref{equ: Delta1}),
\begin{equation}
\label{equ: Delta3}
\left.\begin{aligned}
\Delta_\psi \sim |g| \Big \{ & 2\sin \Big [ {{k^2}\over{4H_m^2}}(\eta-\eta_k) \Big ]+ \\
                      & {{3H}\over{ma}}\cos \Big [ {{k^2}\over{4H_m^2}}(\eta-\eta_k) \Big ] \Big \}.
\end{aligned}
\right.
\end{equation}
Matching $\Delta_{CDM}$ and $\Delta_\psi$ in the limit $k^2/4H_m^2(\eta-\eta_k)\to 0$ but $\eta -\eta_k \gg 1$, we find that $|g|= -18(H_m^2/k^2)\phi_0$ from the sine term of $\Delta_\psi$.  We stress that this result is valid only for long waves, and for short waves $g$ is generally complex.

Note that all modes, except for the very long-wavelength ones that have not yet entered horizon at the radiation-matter equality, must go through this final phase of the radiation era. Equation (\ref{equ: amplitude equation}) is also valid in the matter era, and the solution characteristics deviates from the above description since the self-gravity of $\psi$DM becomes important.  This equation in the matter era has been discussed previously and an analytical solution been obtained \cite{WooChiueh2009}.

\bigskip

(\rmnum{2}) Before mass oscillation

\medskip

Before mass oscillation, the $\psi$DM energy perturbation is always much smaller than the radiation energy perturbation unless particle mass $m > 10^{-28}$ eV, and hence passive evolution can always be a good approximation throughout this regime. In this regime, $\Psi$ is almost a constant, $\Psi^{'}\sim -(m^2a^2/5H)\Psi$, and hence the equation of state of the zero-order field is like that of an inflaton with $P_\psi \sim -\epsilon_\psi$.

\bigskip

(\rmnum{2}a) $ma , k \ll 2H$

\medskip

The super-horizon metric perturbation is again almost a constant, i.e., $\phi \sim \phi_0(1-(1/30)(k^2/H^2)+O(k^4/H^4))$. The perturbed field equation Eq. (\ref{equ: first-order dark matter}) can be approximated by:
\begin{equation}
\label{equ: approximated first order equation dark matter}
\psi^{''}+2H\psi^{'} = -2m^2a^2\Psi\phi,
\end{equation}
to the leading order.  The three terms originally in Eq. (\ref{equ: first-order dark matter}) ignored in Eq. (\ref{equ: approximated first order equation dark matter}) are of order $O(k^2/H^2)$ or $O(m^2a^2/H^2)$ compared with other terms retained.  Similar to Phase (\rmnum{1}a), it turns out that the solution of Eq. (\ref{equ: approximated first order equation dark matter}) also yields a $\Delta_\psi$ that cancels itself to the leading order.

To see the cancellation, we shall find the solution for Eq. (\ref{equ: approximated first order equation dark matter}). The particular integral can be easily obtained by substituting $\psi^{'}\propto (m^2a^2/H)\Psi\phi$ into Eq. (\ref{equ: approximated first order equation dark matter}) and it follows that $\psi_p\sim -(m^2a^2/10H^2)\Psi\phi_0$.  The homogeneous solution decays as $\psi'\sim a^{-2}$ or $\psi'= 0$, and can be ignored.  The energy perturbation for the particular integral becomes $\delta_\psi \sim (-3/25)(ma/H)^2\phi_0$.  On the other hand from Eq. (\ref{equ: momentum perturbation}), $\Psi'\psi \sim (1/25H)(ma/H)^2(m^2a^2\Psi^2/2)\phi_0$, and thus the gauge covariant energy perturbation $\Delta_\psi$ exactly cancels to the leading order of $a^4$.

All corrections of to $\psi_p$, and therefore $\Delta_\psi$, are integer powers of $(k/H)^2$ as only $k^2$ appears in the original equation, Eq.(\ref{equ: first-order dark matter}).  Thus, $\Delta_\psi \sim O((k/H)^2\delta_\psi)\sim O((k/H)^2(ma/H)^2\phi_0)\sim O((ma/H)^2)\Delta_\gamma\propto a^6$.  Here we have again employed the Poisson equation to relate $\Delta_\psi$ to $\Delta_\gamma$.

The $a^6$ rapid growth in this phase is drastically different from the CDM perturbation $(\propto a^2)$, and it is the main cause of the sharp cutoff in the matter power spectrum near a critical $k$ to be discussed in the next section.  Moreover, this $a^6$ rapid growth occurs regardless of whether $k > ma$ or $ma > k$ and modes of all $k$ must go through this initial phase.

\bigskip

(\rmnum{2}b) $ma \ll 2H \ll k$

\medskip

In this regime, the background field is still like an inflaton, where the background energy density is dominated by the field potential energy.  Same as Phase (\rmnum{1}b), the sub-horizon metric perturbation undergoes a damped oscillation, i.e., $\phi\sim \phi_0(H/k)^2\cos(k(\tau-\tau_k)/\sqrt{3})\propto a^{-2}$.  This source drives the perturbed field to also oscillate with the same frequency.  However, unlike Phase (\rmnum{1}a), the driving frequency $k/\sqrt{3}$ is different from the natural frequency $k$ of the $\psi$DM perturbation, resonance is impossible and hence the treatment is different from Phase (\rmnum{1}a).

The source now is dominated by $\Psi'\phi' \sim (k/5\sqrt{3}H)(m^2a^2)\Psi\phi_0\sin(k(\tau-\tau_k)/\sqrt{3})$ and is $\propto a$. Hence the particular integral to the approximate equation
\begin{equation}
\label{equ: approximated amplitude equation}
\psi^{''}+k^2\psi = 4\Psi^{'}\phi^{'}
\end{equation}
is $\psi_p\sim 6\Psi'\phi'/k^2\sim O((m^2a^2/kH)\phi\Psi)\propto a\sin(k(\tau-\tau_k)/\sqrt{3})$.  The energy perturbation $\delta_\psi\sim 2 (\psi'\Psi'-(\Psi')^2\phi)/m^2a^2\Psi^2\sim O((ma/H)^2\phi)\sim O((ma/k)^2\Delta_\gamma)\propto a^2$. The gauge-covariant energy perturbation $\Delta_\psi$ differs from $\delta_\psi$ by a high-order term $O(H/k)$ smaller, and therefore $\Delta_\psi \sim \delta_\psi$, growing as $a^2$ in the oscillation amplitude.

Note that only for sufficiently large $k$ can a mode enter Phase (\rmnum{2}b), and after that it skips Phase (\rmnum{1}a) to go directly to Phase (\rmnum{1}b).   For a mode of low $k$, it initially goes through Phase (\rmnum{2}a) and skips this phase to enter Phase (\rmnum{1}a) and then (\rmnum{1}b).

\section{Evolution of Perturbations in Full Treatment}
\label{sec: Evolution of Perturbations in Full Treatment}

Full treatment of perturbations in the radiation-dominated era must take into account several effects beyond the passive evolution, namely, the decoupling of neutrino after entering horizon, the relative drag between baryons and photons, the non-negligible baryon component in the photon fluid energy density, and the self-gravity of matter.  The last three become significant only near the radiation-matter equality.

\bigskip

(\rmnum{1}) Neutrino Decoupling

\medskip

After entering horizon $k \geq 2H$, photon fluid perturbations oscillate with a constant amplitude indefinitely until the effect of baryon-photon drag sets in.  On the other hand, neutrino perturbations oscillate along with photon fluid perturbations in the first half cycle of oscillation, and after crossing the first null neutrino perturbations are abruptly damped out due to free streaming \cite{MaBertschinger1995}.  We therefore set neutrino density perturbations to zero immediately after its first oscillation null, as an approximation to solve for perturbations of the rest of surviving species.   Thus, we let
\begin{equation}
\label{equ: neutrino Delta}
\Delta_\nu =
\begin{cases}
\Delta_{ph},  &\text{ } \tau \leq \tau_1, \\
0, &\text{ } \tau > \tau_1,
\end{cases}
\end{equation}
where $\Delta_\gamma \equiv \Delta_{ph} + \Delta_\nu$, and $\Delta_\nu$ and $\Delta_{ph}$ are neutrino and photon covariant perturbed energy densities, respectively; $\tau_1$ is the conformal time of the first oscillation null of $\Delta_\gamma$.

\bigskip

(\rmnum{2}) Photon Fluid Equation of State

\medskip

In between the end of radiation-dominant era, $a=a_{eq}$, and the onset of photon-electron decoupling, $a\approx 3a_{eq}$, baryons can be non-negligible in modifying the equation of state parameter for the photon fluid, i.e.,
\begin{equation}
\label{photon fluid EOS}
w_{ph} \equiv {{P_{ph}+P_b}\over{\epsilon_{ph}+\epsilon_b}} = {{a_{eq}}\over{3 \Big ( a_{eq} + a {{\Omega_b}\over{\Omega_m}}{{\Omega_{ph}+\Omega_\nu}\over{\Omega_{ph}}}\Big )}}.
\end{equation}
According to the current values $\Omega_b\approx 0.05$, $\Omega_m = \Omega_{dm}+\Omega_b \approx 0.316$ and $(\Omega_{ph}+\Omega_\nu)/\Omega_{ph} \approx 1.7$ \cite{Planck2016}, the value of $w_{ph}$ approximately equals $20/77$ and $20/111$ when evaluated at the radiation-matter equality and at immediately before photon-electron decoupling, respectively.

In practice, we should not consider a single coupled photon$+$baryon fluid, but should keep track of photon and baryon perturbations separately for the reason given below.  Their respective density equations and coupled momentum equations through Thomson scattering are
\begin{equation}
\label{equ: photon continuity equation}
\delta_{ph}^{'}-{4\over 3}(k^2\theta_{ph}+3\phi^{'})=0,
\end{equation}
\begin{equation}
\label{equ: photon Euler equation}
\theta_{ph}^{'} = - {{\delta P_{ph}}\over{\epsilon_{ph}+P_{ph}}} - \phi + an_e\sigma_T(\theta_b-\theta_{ph}),
\end{equation}
\begin{equation}
\label{equ: baryon continuity equation}
\delta_b^{'}=k^2\theta_b+3\phi^{'},
\end{equation}
and
\begin{equation}
\label{equ: baryon Euler equation}
\theta_b^{'} = - H\theta_b  - \phi - {{4\epsilon_{ph}}\over{3\epsilon_b}}an_e\sigma_T(\theta_b-\theta_{ph}).
\end{equation}
Here $n_e$ is the electron number density and $\sigma_T$ is the Thomson scattering cross section.

\bigskip

(\rmnum{3})  Baryon-Photon Drag

\medskip

Thomson scattering that couples photons and baryons can be characterized by a scattering mean-free-path $l_T\equiv (n_e \sigma_T a)^{-1}$.  After the photon perturbation enters horizon, the wave number $k$ becomes a relevant scale.  Here, we can have a length ratio, $kl_T\propto a^2$.  When $kl_T \ll 1$, the photon perturbation is efficiently coupled to baryons, and photons are well-described by a fluid.  However, there is another length ratio $(k/2H)(kl_T)\propto a^3$ that is larger than $kl_T$ and becomes greater than unity earlier than $kl_T$ does after perturbations entering horizon. The new length scale ratio characterizes when the phase lag between the oscillation of baryon and the oscillation of photon becomes so severe that both photon and baryon perturbations are rapidly damped when $(k/H)(kl_T)>1$.  Physically, photons undergo random walks due to Thomson scattering, and in a Hubble time $H^{-1}$ the random walk distance is $(l_T/H)^{1/2}$.  When this distance is larger than the wavelength, the drag damping can occur \cite{HuSugiyama1996}.  In the wave number range of interest to the present work, $k \sim O(k_c)$, we find $(k/2H)(kl_T)\sim O(1)$ near the radiation-matter equality and the drag damping just begins to take effect.  At this moment, we still have $k l_T \ll 1$ and the photon fluid is a good description and it justifies the photon momentum equation Eq. (\ref{equ: photon Euler equation}).

\bigskip

(\rmnum{4}) Matter Self-Gravity

\medskip

In the full treatment, the source of Poisson equation contains perturbations of all species, to be contrasted with passive evolution where only radiation perturbations are the only source of Poisson equation.  The matter gravity only becomes important near the radiation-matter equality, where the self-gravity prolongs the matter-wave oscillation cycle and slightly modifies the $\psi$DM power spectrum.

\begin{figure}
\includegraphics[scale=0.35, angle=270]{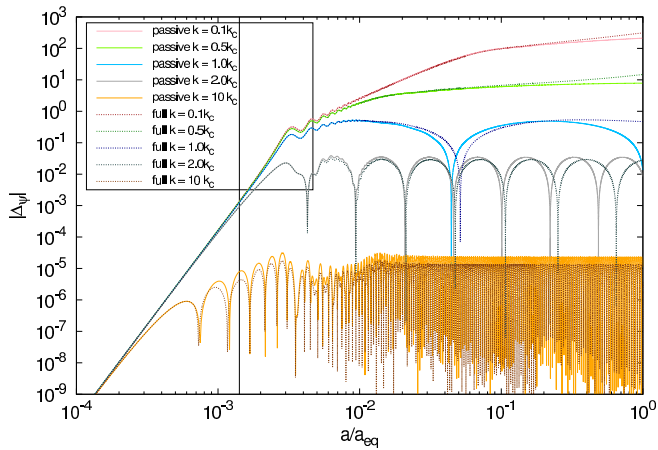}
\caption{Passive evolution and full treatment of gauge covariant $\psi$DM energy density perturbations for several different k's.   The full treatments (dotted lines) deviate from the passive evolution (solid lines) only near the epoch of radiation-matter equality except for very high-$k$ modes, as explained in the text.}
\label{fig: compare_passive_active_free_evolution_more_modes}
\end{figure}

A 4-th order Runge-Kutta scheme is adopted for integrating the background equations (Eqs. (\ref{equ: zeroth-order dark matter}) and Friedmann equations described in Appendix (\ref{app: full treatment evolution}), i.e., Eqs. (\ref{equ: Friedmann equations 1 with baryon and neutrino}) and (\ref{equ: Friedmann equations 2 with baryon and neutrino}), with equations of states) and the perturbed equations (Eqs. (\ref{equ: first-order dark matter}), (\ref{equ: photon continuity equation}), (\ref{equ: photon Euler equation}), (\ref{equ: baryon continuity equation}), (\ref{equ: baryon Euler equation}) and perturbed Einstein equations described in Appendix (\ref{app: full treatment evolution}), i.e., Eqs. (\ref{equ: perturbed Einstein equation 1 wih baryon and neutrino}) and (\ref{equ: perturbed Einstein equation 2 wih baryon and neutrino}), with the neutrino contribution given in Eqs. (\ref{equ: delta_nu}) and (\ref{equ: theta_nu})). Plotted with dotted lines in Fig. (\ref{fig: compare_passive_active_free_evolution_more_modes}) are the numerical solutions of the covariant energy perturbations $\Delta_\psi$ for full treatment as functions of the scaling factor $a/a_{eq}$ for particle mass $m=10^{-22}$eV. We have chosen modes with $k \gg k_c$ and $k \ll k_c$ to illustrate the difference.  The vertical solid line marks the onset of mass oscillations ($a = a_m$). The three asymptotic phases discussed in the last section are clearly shown in Fig. (\ref{fig: compare_passive_active_free_evolution_more_modes}) for the low-$k$ mode and the other three for higher-$k$ mode.    Solutions of passive evolution are also plotted as solid lines in Fig. (\ref{fig: compare_passive_active_free_evolution_more_modes}) for comparison.  One notices that the full treatment and the passive evolution deviate from each other only near the radiation-matter equality, indicative of the slowing down of matter-wave oscillation due to the matter self-gravity.   It is not surprising to find that treating neutrinos as a fluid for passive evolution does not differ from a more sophisticated approximation for neutrinos in the full treatment, as long as $k$ is smaller than $k_c$.  This is due to that sub-horizon perturbations of these low-$k$ modes are hardly coupled to metric perturbations (Phase(\rmnum{1}b)).  Noticeable differences occur only for $k >k_c$ since these modes respond to photon oscillations driven by metric perturbations after entering horizon (Phase(\rmnum{2}b)), and a decrement in the metric perturbation due to the vanishing neutrino contribution decreases the oscillation amplitude.  These high-$k$ modes, however, have vanishingly small power at the radiation-matter equality and hence the differences are insignificant for practical purposes.

In Appendix (\ref{app: full treatment evolution}), we show in detail how the photon fluid and the baryon fluid, coupled by the relative drag, evolve in the full treatment. To summarize, passive evolution of $\psi$DM captures most essential physics in the evolution of full treatment. Hence we shall continue to take advantage of the understanding gained from passive evolution for the following discussions.

\section{Critical Mode, Matter Power Spectrum and Sub-Horizon Dynamics}
\label{sec: Critical mode and Sub-Horizon Dynamics}

We saw in the previous discussions that high-$k$ modes go through Phases (\rmnum{2}a), (\rmnum{2}b) and (\rmnum{1}b), and low-$k$ modes go through Phase (\rmnum{2}a), (\rmnum{1}a) and (\rmnum{1}b).  In between the two, there exists a critical $k_c$ mode crudely defined as $k_c=ma_m=2H_m = (2H_m ma_m)^{1/2}=(2H ma)^{1/2}$, a redshift-independent wave number, that characterizes a transition.  This particular mode enters horizon at the moment when the mass oscillation starts, and so the mode can be viewed as having a wavelength equal to the Compton wavelength when $\psi$DM is crystalized into a real particle.  For modes of $k<k_c$, they have wavelengths greater than the Compton length ever since the particle becomes real; for modes of $k > k_c$, there is a finite period after the particle becomes real where their wavelengths are smaller than the Compton length. One expects that structures smaller than the Compton length can hardly exist and hence must be suppressed. This is indeed what we saw in the drastically different dynamics for modes of $k< k_c$ and $k > k_c$.

\begin{figure}
\includegraphics[scale=0.35, angle=270]{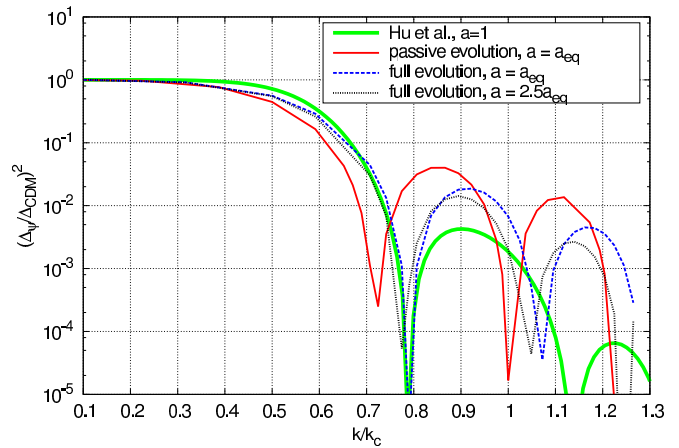}
\caption{The $\psi$DM transfer function relative to CDM evaluated at $a_{eq}$ and $2.5a_{eq}$. The horizontal axis is the wavenumber normalized to $k_c$. The solid line is for the passive evolution and the dashed line for the full treatment evaluated at $a_{eq}$. The dotted line is the full treatment evaluated at $2.5 a_{eq}$ where the plasma is still fully ionized.  The particle mass is the same with Fig. (\ref{fig: compare_passive_active_free_evolution}).  We see the passive evolution and the full treatment makes very little difference for $k < 0.4k_c$, and both follow the CDM power spectrum closely.  But for $k > 0.4k_c$, the two begin to have noticeable differences, due primarily to slightly different frequencies in the matter-wave oscillation. The transfer function for the full treatment at $a=a_{eq}$ also has little difference from that at $a=2.5 a_{eq}$ for $k < k_c$.  Above $k_c$, the wave function oscillates during this period.  The bold line is the analytical fitting formula given by \cite{Hu2000} evaluated at the present, where further short-wave suppression well after $a=a_{eq}$ is clearly seen.}
\label{fig: transfer_function_compare_free_passive_active}
\end{figure}

Plotted in Fig. (\ref{fig: transfer_function_compare_free_passive_active}) are the $\psi$DM transfer function in reference to CDM, $|\Delta_\psi|^2/|\Delta_{CDM}|^2$, evaluated at $a_{eq}$ as functions of $k$ with particle mass, $m=10^{-22}$eV. The solid line denotes the passive evolution and the dashed line the full treatment. Comparison of the two curves shows minor ($<10\%$) differences for $k < k_c/2$, demonstrating the passive evolution is a good approximation for the entire radiation-dominant era for these long waves.  For $k > k_c/2$, our discussion about the mode suppression still holds and details are given below.  In addition, Fig. (\ref{fig: transfer_function_compare_free_passive_active}) exhibits substantially different oscillation phases for the two (passive versus full treatment) evolutionary paths. This is caused by the self-gravity in full treatment that lowers the matter-wave oscillation frequency. But the self-gravity comes to play late in the evolution, and hence little affects our dynamical picture of Phase (\rmnum{1}b).

In Fig. (\ref{fig: transfer_function_compare_free_passive_active}), we also provide the power spectrum evaluated at $2.5a_{eq}$, serving as the initial condition for any calculation starting in the matter-dominated era.

Below, we shall explain how the short-wave suppression occurs more quantitatively. All modes first follow the rapid $a^6$ growth of phase (\rmnum{2}a).  Modes with $k < k_c$ continue this growth until the mass oscillation starts at $a=a_m$, and then they follow the CDM growth before entering horizon.  Modes with $k > k_c$ exit the rapid growth of Phase (\rmnum{2}a) early on when becoming sub-horizon and enter the slower $a^2\cos(k(\tau-\tau_k))$ growth of Phase (\rmnum{2}b) before mass oscillation, and after mass oscillation, they enter Phase (\rmnum{1}b) directly without going through the CDM phase, Phase (\rmnum{1}a).   Hence, these high-$k$ modes never have a chance to resemble CDM perturbations. The amplitudes differ roughly by $(k_c/k)^4$ for modes with wave numbers few times above $k_c$ in comparison with the critical mode at the onset of mass oscillation. Thus, the sharp spectral transition has already been established when mass oscillation starts and real particles of finite mass $m$ are crystalized.

Additional suppression also occurs in Phase (\rmnum{1}b), which mostly affects modes near $k_c$. This suppression has nothing to do with the Compton length effect mentioned earlier, but something to do with the Jeans length.   While the CDM mode still grows mildly as $\ln(a/a_k)$, the $\psi$DM mode in Phase (\rmnum{1}b) oscillates with a constant amplitude, as $(2Hma/k^2)\sin[(k^2/2Hma)\ln(a/a_k)]$, leading to a $k^{-2}$ suppression for shorter waves.  This Jeans suppression in the radiation-dominant regime is an extension of the more familiar counterpart in the matter-dominant regime.

Despite the formula given in the last paragraph for $\psi$DM modes was derived for $k \ll k_c$, it still roughly holds for $k\sim k_c$.  Take the critical mode with particle mass $m =10^{-22}$ eV as an example.  The mass oscillation begins around the photon temperature $T_\gamma \sim 0.5 $ keV, occurring $1/500$ times smaller in $a$ than that at the radiation-matter equality with $T_\gamma\sim 1$ eV, and hence the oscillation amplitude of the critical mode is $(\ln(a_{eq}/a_m))^{-1}$ smaller than that of the CDM mode of the same $k$. We thus have an oscillation amplitude suppression factor of about $6.2$, pretty consistent with Fig.(\ref{fig: transfer_function_compare_free_passive_active}). In general, the oscillation amplitude suppression factor is approximately $(\ln(500k/k_c))^{-1}(k_c/k)^2$ around $k_c$. This formula is still valid for a different particle mass $m$. For that purpose one simply replaces $500$ by $a_{eq}/a_m$.

Accompanying the suppression of the high-$k$ power is the oscillation pattern for shorter waves in the power spectrum.  There are two types of oscillations separately exhibited by modes of $k\gg k_c$ and $k\sim k_c$.  For longer waves, the oscillation pattern arises primarily from the matter-wave oscillation and for shorter waves from the photon fluid oscillation followed by the matter-wave oscillation.  Although the oscillation pattern in the power spectrum for shorter waves is complicated, due to the mixture of two oscillations of different frequencies, the one for longer waves is predictable. The phase of the latter oscillation is given by $(k/k_c)^2\ln(a_{eq}/a_k)$ and hence the oscillation frequency in the power spectrum is $[(k/k_c)^2(\ln(a_{eq}/a_m)+\ln(k_c/k))]^2$.  The transition between the two oscillation patterns for longer waves and for shorter waves is found roughly at $k=2.5k_c$ for $m\sim 10^{-22}$eV.  (The power spectrum shown in Fig. (\ref{fig: transfer_function_compare_free_passive_active}) is hence not extended to cover such a short-wave regime.)

One can conveniently define $k_{1/2}$ as the wavenumber for which the modal power is suppressed by $1/2$ relative to CDM.  Taking into account the phase and amplitude of the matter-wave oscillation, we find that $k_{1/2}/k_c \sim (\alpha/\ln(m/m_{28}))^{1/2}$ for any $m\gg m_{28}$ and $\alpha \approx 1.3$ for passive evolution and $\alpha \approx 1.7$ for active evolution, where $m_{28}\equiv 10^{-28}$eV. The mass dependence arises from that the onset of mass oscillations for a more massive particle occurs earlier $(a_m\propto m^{-1/2})$, and therefore it has a more ample time to execute matter-wave oscillations before the radiation-matter equality.

Interestingly the Jeans wave number $k_J$ in the matter dominated regime, for which modes with $k < k_J$ grow like CDM and modes with $k > k_J$ oscillate with constant amplitudes, is also $\sqrt{2Hma}$.  Moreover in the matter-dominated regime, $H\propto a^{-1/2}$ and hence the Jeans length $k_J\propto a^{1/4}$, mildly evolving toward shorter wavelengths.  This explains why $k_J$ is always larger than, but on the same order of, $k_c$. The oscillation pattern in the power spectrum near $k_c$ is therefore frozen in the matter-dominant era and grows self-similarly as $a^2$.

\begin{figure}
\includegraphics[scale=0.35, angle=270]{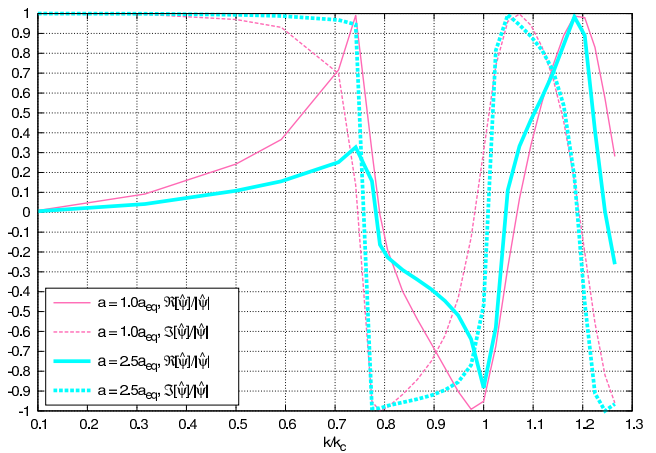}
\caption{Phases of wave functions, $\Re[\hat\psi]/|\hat\psi|$ and $\Im[\hat\psi]/|\hat\psi|$, evaluated at $a=a_{eq}$ and $a=2.5 a_{eq}$ as functions of $k$, and obtained by the full treatment. This plot also allows one to narrow down the value of Jeans length during this period as explained in the main text.}
\label{fig: ampltude_psi_hat}
\end{figure}

It is relevant to also compute the phase $S$ of $\hat\psi$, which, along with the power spectrum, can determine $\psi$ used for simulations in the matter-dominant regime as the initial condition.  The power spectrum $\langle (\Re[\hat\psi])^2\rangle=\langle|\hat\psi|^2\rangle\cos^2(S)$ only provides a partial condition for this purpose.  When $\Re[\hat\psi]/|\hat\psi|=\cos(S)$ and $\Im[\hat\psi]/|\hat\psi|=\sin(S)$ are available, one can construct $\exp(iS)$ for each $k$ mode.  The amplitude $|\hat\psi|$, being a half-Gaussian random number, can be then fixed by constructing an exponential random number $|\hat\psi|^2\cos(S)^2$ and letting its average equal to the power spectrum. Figure (\ref{fig: ampltude_psi_hat}) shows the ratio $\Re[\hat\psi]/|\hat\psi|$ as functions of $k$ at the matter-radiation equality ($a=a_{eq}$) and at $a=2.5a_{eq}$ using the full treatment. The first thing brings to notice is $\Im[\hat\psi] \gg \Re[\hat\psi]$ for long waves.  This feature is apparent from the discussions of Phase (\rmnum{1}b).  Secondly, $\Im[\hat\psi]$ grows substantially faster than $\Re[\hat\psi]$ in between $a=a_{eq}$ and $a=2.5a_{eq}$ due to the acceleration of self-gravity, and the ratio $\Re[\hat\psi]/\Im[\hat\psi] \to 0$ from above for most $k < k_c$ modes.  These modes are purely growing modes.  For $k > k_c$, the wave function oscillates with a time-varying frequency, and in the range $k_c < k < 1.3 k_c$ plotted here, the oscillation frequency is nearly zero, indicative of that the wavenumber is close to the Jeans wavenumber $k_J$.

\section{Adiabatic Perturbations for Superhorizon Modes}
\label{sec: Adiabatic Perturbations}

In the multi-fluid model, the condition for adiabatic perturbation is Eq. (\ref{equ: adiabatic condition}).  Here $\langle w\rangle$ is the short-time averaged equation of state parameter $\langle w\rangle=\langle P/\epsilon\rangle$.

Before the mass oscillation in Phase (\rmnum{2}a), there is no fast mass oscillation and the equation of state for $\psi$DM is
\begin{equation}
\label{equ: equation of state}
1+\langle w_{\psi}\rangle \sim 2\Big ( {{ma}\over{5H}} \Big )^2 \to 0,
\end{equation}
and in this phase $\psi$DM behaves similar to the dark energy.  In Appendix (\ref{app: passive evolution}) we see $\delta_\gamma\sim -2\phi$ in this phase, approximately a constant. Hence the adiabatic condition yields $\delta_\psi = -(3/25)(m^2a^2/H^2)\phi$, the same as we found by solving the approximate dynamical equation.

After mass oscillation in Phase (\rmnum{1}a), we have used the adiabatic condition to pin down the value of the super-horizon energy perturbation, which involves an undetermined constant.  In fact, one may question whether Eq. (\ref{equ: adiabatic condition}) in Phase (\rmnum{1}a) really needs to take a short-time average.  This question can be answered by examining Eq. (\ref{equ: deltapsi}) and the fact that $1+w_\psi\propto(\Psi')^2\propto \sin^2(mt + \alpha)$.  Since $\delta_\gamma/(1+w_\gamma)$ has no mass oscillation but both $\delta_\psi$ and $w_\psi$ do, one should examine whether the respective oscillations of $\delta_\psi$ and $1+w_\psi$ cancel.  We had before $\Re[\hat\psi] \sim -(3/4)\phi_0$, and $\Im[\hat\psi]_p=-(ma/2H)\phi_0$.  Upon substituting them into Eq. (\ref{equ: deltapsi}), we find  $\delta_\psi$ is also proportional to $\sin^2(mt+\alpha)$, and $\delta_\psi/(1+w_\psi)$ indeed has no oscillation.  Therefore the adiabatic condition holds instantaneously without the need to take a short-time averages over $\delta_\psi$ and $w_\psi$, even though the perturbation has fast oscillations.

Next, one may wonder why the long-wavelength adiabatic perturbations of $\psi$DM and CDM should track each other so well after the mass oscillation.  A rough answer is that since they are both adiabatic and when the equations of state are the same, the two must be identical when super-horizon.  As we have seen, the average equation of state parameter $<w_\psi>=0$ in Phase (\rmnum{1}a), the same as that of CDM, which justifies the above statement.  Further considerations show that once the mode enters Phase (\rmnum{1}b), we also have the same fluid description for $\psi$DM as that for CDM in the long wave limit, since $\delta P_\psi =O((k/ma)^2 \delta\epsilon_\psi)\to 0$ when $k\to 0$.  Therefore, for sufficiently long waves, $\psi$DM and CDM become almost identical after mass oscillation, despite the two are very different before mass oscillation.

\section{Axion Model}
\label{sec: Axion Model}

Axion has a field potential $V=(fm)^2(1-\cos(\Psi/f))$, where there appears a new energy scale, the axion decay constant $f$, believed to be close to the GUT scale or the Planck scale \cite{ChiueharXiv2014, HuiarXiv2016}.  In the limit $\Psi/f\to 0$, the axion potential is reduced to the free-particle harmonic potential discussed previously.  Normally without fine tuning, it is expected that the initial axion angle $\theta(\equiv\Psi/f)$ is of order unity at the very early epoch.  (Here we have taken a simple axion model for which the field potential has been established long before the mass oscillation.)

Unlike the free particle model, the zero-order field obeys a nonlinear equation:
\begin{equation}
\label{equ: zeroth order axion model equation}
\theta^{''}+2H\theta^{'}+m^2a^2\sin(\theta) =0.
\end{equation}
Similar to Eq. (\ref{equ: zeroth-order dark matter}), $\theta$ can oscillate, but initially executing anharmonic oscillation.  Subject to the Hubble friction, $\theta$ declines in amplitude after a couple of oscillations where $\sin(\theta) \to \theta$, and Eq. (\ref{equ: zeroth order axion model equation}) becomes Eq. (\ref{equ: zeroth-order dark matter}).  Thus, the axion model is a straightforward extension of the free-particle model, and can address the initial phase-lock problem of $\psi$DM explained in Sec. (\ref{sec: Introduction}).

The perturbed field now obeys
\begin{equation}
\label{equ: first order axion model equation}
\delta\theta^{''}+2H\delta\theta^{'} + k^2\delta\theta +m^2a^2\cos(\theta)\delta\theta = 4\theta^{'}\phi^{'}-2m^2a^2\sin(\theta)\phi,
\end{equation}
which recovers Eq. (\ref{equ: first-order dark matter}) when $\theta << 1$.  The only difference of the axion model from the free-particle models occurs near the $ma=2H$ transition, and the one additional free parameter for the axion model is the initial $\theta$, which we denote as $\theta_0$.

\begin{figure}
\includegraphics[scale=0.35]{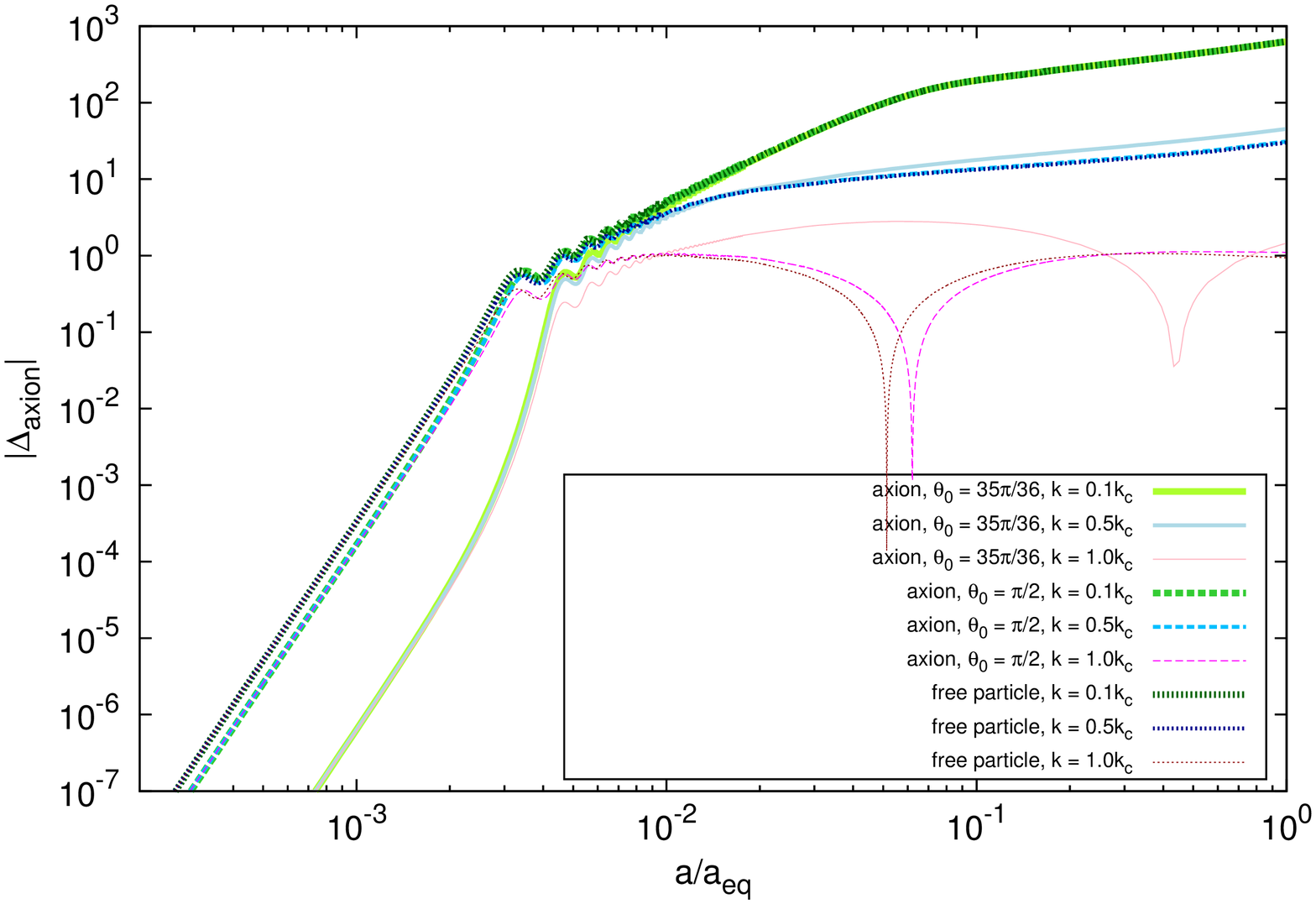}
\caption{Full treatments of the axion model with three different initial angles $\theta_0$. The dotted line represents $\theta_0 << 1$ (free-particle model), the dashed line $\theta_0=\pi/2$ and the solid line $\theta_0=35\pi/36$.  We also plot three different wavenumbers, $k << k_c$, $k=0.5 k_c$ and $k= k_c$, for the above three $\theta_0$.  For $k << k_c$, the initial angle $\theta_0$ does not make any difference in covariant energy density perturbations $\Delta_{axion}$.  For $k = 0.5 k_c$, some differences appear, particularly for the $\theta_0=35\pi/36$ case, which has a higher amplitude.  For $k=k_c$, the $\theta_0=35\pi/36$ case becomes very different from the other two cases.   For all three $k$'s, the $\theta_0 <<1$ and $\theta_0=\pi/2$ cases have similar $\Delta_{axion}$, indicative of that $\Delta_{axion}$ is insensitive to $\theta_0$, unless $\theta_0$ assumes an extreme value very close to $\pi$.}
\label{fig: compare_active_axion_middle_top}
\end{figure}

In contrast to the free-particle model, the axion model has no analytical solutions for the zero-order field, and hence the analytical solution to the first-order field is not attainable even for passive evolution. We replace Eqs. (\ref{equ: zeroth-order dark matter}) and (\ref{equ: first-order dark matter}) by Eq. (\ref{equ: zeroth order axion model equation}) and Eq. (\ref{equ: first order axion model equation}) and numerically integrate the coupled equations of the axion model as we did for the full treatment of the free-particle model.  The results are plotted in Fig. (\ref{fig: compare_active_axion_middle_top}) for $\theta_0 = \pi/16, \pi/2, 35\pi/36$, where the $\theta_0 =\pi/16$ case is essentially the free-particle model.

\begin{figure}
\includegraphics[scale=0.35]{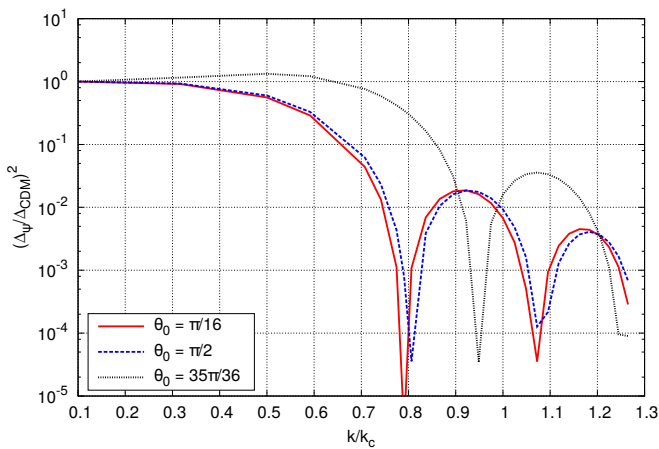}
\caption{Axion perturbation transfer functions relative to CDM evaluated at $a_{eq}$.  Full treatment results of three different axion initial angles are presented.  Again, the case with $\theta_0=\pi/2$ (dashed line) is almost the same as with $\theta_0=\pi/16$ (thin solid line).  On the other hand, for the extreme case with $\theta_0=35\pi/36$ (dotted line), $k_{1/2}$ extends to substantially higher $k$ with its amplitude slightly exceeding $1$.}
\label{fig: compare_axion_free_particle_transfer_function_a_eq}
\end{figure}

Perturbations of the axion model also roughly have the four asymptotic solutions corresponding to the four asymptotic phases of the free-particle model.  This is actually not surprising since the major difference between the two models is just during the time near the onset of mass oscillation; for a difference in such a short time, the asymptotic phases are little affected. Generally speaking, the initial anharmonic oscillation of the axion model has longer periods than the harmonic oscillation of the free-particle model and would lead to some phase delay of mass oscillation.  But solutions at $a \gg a_m$ turn out not to be sensitive to such a moderate phase shift.  This is clearly shown in the power spectrum of Fig. (\ref{fig: compare_axion_free_particle_transfer_function_a_eq}) by a comparison of solutions of the free-particle model and the axion model of a moderate initial angle $\theta_0 = \pi/2$. The difference is at most $10\%$ for $k \leq k_{1/2}$.  Only in the extreme case when $\theta_0 \to \pi$, where the onset of mass oscillation is significantly delayed causing the axion's first cycle of matter-wave oscillation discussed in Phase (\rmnum{1}b) to take a substantially longer period than that of the free particle, can it make an appreciable difference.  The longer matter-wave oscillation cycle changes the phase evaluated at $a=a_{eq}$ and can yield a higher $k_{1/2}$, as a comparison of $k=k_c$ cases for three different $\theta_0$'s hints in Fig. (\ref{fig: compare_active_axion_middle_top}).   This feature makes the axion of extreme $\theta_0$ appear as if the particle had a higher mass assuming it is free particle that follows $k_{1/2}\sim O(k_c) \propto m^{1/2}$.  Surprisingly, the axion model of extreme initial angle can also produce a slightly higher power than the CDM model for the extreme $\theta_0\to \pi$ case, which has never been so for the free-particle model, and the amplitude excess occurs near $k= k_c/2$ for $\theta_0=35\pi/36$.   For the fiducial particle mass $10^{-22}$ eV, the power excess is in the wavelength range $>1$Mpc, which encloses about  $10^{10}- 10^{11} M_\odot$ of dark matter, a typical mass of first galaxies, and may have an impact on the first-galaxy formation.

How do we understanding the two new features, higher $k_{1/2}$ and amplitude excess near $k_c/2$ of the extreme axion model?   Compared with the free-particle model, the axion perturbation $\delta\theta$ has much smaller sources initially to drive the growth when $\theta$ is close to $\pi$, as its potential gradient is almost zero. This is why $\Delta_{axion} \ll \Delta_\psi$ in Phase (\rmnum{2}a).  In the beginning, this small source is balanced by both "inertia" and "friction", similar to the free-particle case, and $\Delta_{axion}\propto a^6$.  Shortly after, $\theta$ is rolling down the potential hill, the field force much exceeds the friction, and the perturbation grows more rapidly than $a^{6}$.  This is almost an exponential growth, but it only lasts briefly for a fraction of the first mass oscillation cycle. In the first few mass oscillation cycles after the rapid growth, the axion perturbation grows roughly as $a^3$, as contrasted to the $a^2$ growth of the free-particle perturbation, and does the final catch-up with the free particle perturbation.  The subsequent evolution of the axion perturbation after entering horizon pursues a slightly different evolutionary path from the free-particle perturbation, albeit both types of perturbations at this point satisfy almost the same Eq. (\ref{equ: amplitude equation 1}) except for a small nonlinear correction in the axion case.

To assess this problem properly, we must keep the nonlinearity, albeit small, of the background $\theta$. The nonlinear mass oscillation causes nonlinear frequency shifts in both $\theta$ and $\delta\theta$.  But two frequency shifts are slightly different, leading to additional detuning between the response and the driving source, similar to the role of $k$. (See the first paragraph of Phase (\rmnum{1}a).)  Since the matter-wave oscillation frequency is the detune frequency, the nonlinear effect changes the matter-wave oscillation frequency.  The nonlinearity decreases with time, and hence the earlier the mode enters horizon, the more prominent this effect is. It explains why the very-long wavelength extreme axion perturbation, entering horizon very late, has no effect and is almost identical to the free-particle perturbation.

Moreover, the detune frequency yielded by the nonlinear oscillation of $\theta_0$ has two components, a quasi-static component and a rapidly oscillating component.  It turns out the rapidly oscillating component drives the perturbed field into a parametrically unstable state and produces the power excess.

The above mechanisms qualitatively explain the most important two features of the extreme axion model. Given the implications that the extreme axion model can yield a higher particle mass effect in the spectral cutoff and can also produce a higher power than CDM in the wavelength range that yields first galaxies, this subject warrants a quantitative analysis, and we report it in a separate paper \cite{ZhangChiueharXiv2017}.

\section{Conclusions}
\label{sec: conclusions}

In this paper, why the matter power spectrum at the end of radiation era exhibits a CDM-like spectrum for long waves and a highly suppressed spectrum for short waves and why the transition of these two types of spectra is sharp are fully explored.  Physically the sharp power suppression is due to the smearing of fluctuations below the Compton length.  Dynamically it is due to short waves prematurely entering horizon before mass oscillations, so that quantum pressure becomes so effective as to slow down the otherwise very rapid growth which long waves entertain.

To summarize, the following issues are addressed and results obtained:

(1) We elucidate four asymptotic phases with passive evolution--- before mass oscillations, after mass oscillations, super-horizon wavelengths and sub-horizon wavelengths.  The transition occurs at the intersection of the four asymptotic phases, i.e., at the onset of mass oscillations for the wavelength just beginning to enter the horizon, i.e., $k = 2H = ma$.  It defines a critical wavenumber $k_c$ which coincides the transition of the two distinct behaviors in the $\psi$DM spectrum.  This $k_c$ is the predecessor in the radiation-dominant era of the Jeans wavenumber $k_J$ in the matter-dominant era.

(2) We demonstrate passive evolution of $\psi$DM to be a good approximation of the full treatment, and we also address the details of the full treatment, including collisionless neutrinos and Thomson scattering coupled photons and baryons, through which numerical solutions are obtained.

(3) As a bonus of our analyses, we show that the adiabatic condition for super-horizon perturbations holds instantaneously even when the field potential produces fast mass oscillations.

(4) The phase and amplitude of the complex $\hat\psi$ are important quantities for any $\psi$DM simulation to start the initial condition faithfully in the matter dominant era.  For this reason, we extend our numerical solutions into the matter-dominant era and compute the phase of $\hat\psi$ and the power spectrum beyond the radiation-matter equality.

(5) We also consider perturbations of the axion model, a plausible extension of the free-particle model of $\psi$DM.  It is found that the evolution of axion perturbations is generally almost identical to that of free-particle perturbations, except for the extreme case where the initial value of the axion angle is near the field potential top, yielding a substantially longer frequency of matter wave oscillation and shifting the transition wavenumber to a higher value. The upward shift in the transition wavenumber is equivalent to having a higher particle mass for the free particle model.  This extreme axion model also produces a higher power than CDM in a wavelength range about half of the critical wavenumber.  For particle mass $\sim 10^{-22}$eV, the enclosed mass of these wavelengths is the typical mass of first galaxies, which may have impact to first galaxies at the cosmological redshift $z\sim 10$, discovered by recent observations \cite{Bouwens2011, Bouwens2012, Zheng2012, Coe2013, Oesch2014, Zitrin2014, Bouwens2015, Bernard2016}.

This work is supported in part by MOST of Taiwan under the grant MOST 103-2112-M-002-020-MY3.

\bibliography{Reference}

\begin{thebibliography}{65}
\expandafter\ifx\csname natexlab\endcsname\relax\def\natexlab#1{#1}\fi
\expandafter\ifx\csname bibnamefont\endcsname\relax
  \def\bibnamefont#1{#1}\fi
\expandafter\ifx\csname bibfnamefont\endcsname\relax
  \def\bibfnamefont#1{#1}\fi
\expandafter\ifx\csname citenamefont\endcsname\relax
  \def\citenamefont#1{#1}\fi
\expandafter\ifx\csname url\endcsname\relax
  \def\url#1{\texttt{#1}}\fi
\expandafter\ifx\csname urlprefix\endcsname\relax\def\urlprefix{URL }\fi
\providecommand{\bibinfo}[2]{#2}
\providecommand{\eprint}[2][]{\url{#2}}

\bibitem[{\citenamefont{{Hu} et~al.}(2000)\citenamefont{{Hu}, {Barkana}, and {Gruzinov}}}]{Hu2000}
  \bibinfo{author}{\bibfnamefont{W.}~\bibnamefont{{Hu}}},
  \bibinfo{author}{\bibfnamefont{R.}~\bibnamefont{{Barkana}}}, \bibnamefont{and}
  \bibinfo{author}{\bibfnamefont{A.}~\bibnamefont{{Gruzinov}}},
  \bibinfo{journal}{Phys. Rev. Lett.} \textbf{\bibinfo{volume}{85}},
  \bibinfo{pages}{1158} (\bibinfo{year}{2000}).

\bibitem[{\citenamefont{Schive et~al.}(2014)\citenamefont{{Schive}, {Chiueh}, and {Broadhurst}}}]{Schive2014}
  \bibinfo{author}{\bibfnamefont{H.-Y.} \bibnamefont{Schive}},
  \bibinfo{author}{\bibfnamefont{T.}~\bibnamefont{Chiueh}}, \bibnamefont{and}
  \bibinfo{author}{\bibfnamefont{T.}~\bibnamefont{Broadhurst}},
  \bibinfo{journal}{Nature Phys.} \textbf{\bibinfo{volume}{10}},
  \bibinfo{pages}{496} (\bibinfo{year}{2014}).

\bibitem[{\citenamefont{{Marsh} and {Pop}}(2015)}]{MP2015}
  \bibinfo{author}{\bibfnamefont{D.~J.~E.} \bibnamefont{{Marsh}}}, \bibnamefont{and}
  \bibinfo{author}{\bibfnamefont{A.~R.} \bibnamefont{{Pop}}}, 
  \bibinfo{journal}{Mon. Not. R. Astron. Soc.} \textbf{\bibinfo{volume}{451}}, 
  \bibinfo{pages}{2479} (\bibinfo{year}{2015}).

\bibitem[{\citenamefont{{Calabrese} and {Spergel}}(2016)}]{CS2016}
  \bibinfo{author}{\bibfnamefont{E.} \bibnamefont{{Calabrese}}}, \bibnamefont{and}
  \bibinfo{author}{\bibfnamefont{D.~N.} \bibnamefont{{Spergel}}}, 
  \bibinfo{journal}{Mon. Not. R. Astron. Soc.} \textbf{\bibinfo{volume}{460}}, 
  \bibinfo{pages}{4397} (\bibinfo{year}{2016}).

\bibitem[{\citenamefont{Hui et~al.}(2016)\citenamefont{{Hui}, {Ostriker}, {Tremaine},and {Witten}}}]{HuiarXiv2016}
  \bibinfo{author}{\bibfnamefont{L.} \bibnamefont{Hui}},
  \bibinfo{author}{\bibfnamefont{J.~P.} \bibnamefont{Ostriker}},
  \bibinfo{author}{\bibfnamefont{S.}~\bibnamefont{Tremaine}}, \bibnamefont{and}
  \bibinfo{author}{\bibfnamefont{E.}~\bibnamefont{Witten}},
  {arXiv:1610.08297} (\bibinfo{year}{2016}). 

\bibitem[{\citenamefont{{Lora} and {Maga{\~n}a}}(2014)}]{LM2014}
  \bibinfo{author}{\bibfnamefont{V.}~\bibnamefont{{Lora}}} \bibnamefont{and}
  \bibinfo{author}{\bibfnamefont{J.}~\bibnamefont{{Maga{\~n}a}}},
  \bibinfo{journal}{J. Cosmology Astropart. Phys.} \textbf{\bibinfo{volume}{09}},
  \bibinfo{pages}{011} (\bibinfo{year}{2014}).

\bibitem[{\citenamefont{{Moore}}(1994)}]{Moore1994}
  \bibinfo{author}{\bibfnamefont{B.}~\bibnamefont{{Moore}}},
  \bibinfo{journal}{Nature} \textbf{\bibinfo{volume}{370}},
  \bibinfo{pages}{629} (\bibinfo{year}{1994}).

\bibitem[{\citenamefont{{Goerdt} et~al.}(2006)\citenamefont{{Goerdt}, {Moore}, {Read}, {Stadel}, and {Zemp}}}]{Goerdt2006}
  \bibinfo{author}{\bibfnamefont{T.}~\bibnamefont{{Goerdt}}},
  \bibinfo{author}{\bibfnamefont{B.}~\bibnamefont{{Moore}}},
  \bibinfo{author}{\bibfnamefont{J.~I.}~\bibnamefont{{Read}}},
  \bibinfo{author}{\bibfnamefont{J.}~\bibnamefont{{Stadel}}}, \bibnamefont{and}
  \bibinfo{author}{\bibfnamefont{M.}~\bibnamefont{{Zemp}}},
  \bibinfo{journal}{Mon. Not. R. Astron. Soc.} \textbf{\bibinfo{volume}{368}},
  \bibinfo{pages}{1073} (\bibinfo{year}{2006}).

\bibitem[{\citenamefont{{Gilmore} et~al.}(2007)\citenamefont{{Gilmore}, {Wilkinson}, {Wyse}, {Kleyna}, {Koch}, {Evans}, and {Grebel}}}]{Gilmore2007}
  \bibinfo{author}{\bibfnamefont{G.}~\bibnamefont{{Gilmore}}},
  \bibinfo{author}{\bibfnamefont{M.~I.} \bibnamefont{{Wilkinson}}},
  \bibinfo{author}{\bibfnamefont{R.~F.~G.} \bibnamefont{{Wyse}}},
  \bibinfo{author}{\bibfnamefont{J.~T.} \bibnamefont{{Kleyna}}},
  \bibinfo{author}{\bibfnamefont{A.}~\bibnamefont{{Koch}}},
  \bibinfo{author}{\bibfnamefont{N.~W.} \bibnamefont{{Evans}}}, \bibnamefont{and} 
  \bibinfo{author}{\bibfnamefont{E.~K.} \bibnamefont{{Grebel}}}, 
  \bibinfo{journal}{Astrophys. J.} \textbf{\bibinfo{volume}{663}}, 
  \bibinfo{pages}{948} (\bibinfo{year}{2007}).

\bibitem[{\citenamefont{{Walker} and {Pe{\~n}arrubia}}(2011)}]{WP2011}
  \bibinfo{author}{\bibfnamefont{M.~G.} \bibnamefont{{Walker}}} \bibnamefont{and}
  \bibinfo{author}{\bibfnamefont{J.}~\bibnamefont{{Pe{\~n}arrubia}}},
  \bibinfo{journal}{Astrophys. J.} \textbf{\bibinfo{volume}{742}},
  \bibinfo{pages}{20} (\bibinfo{year}{2011}).

\bibitem[{\citenamefont{{Amorisco} et~al.}(2013)\citenamefont{{Amorisco}, {Agnello}, and {Evans}}}]{AAE2013}
  \bibinfo{author}{\bibfnamefont{N.~C.} \bibnamefont{{Amorisco}}},
  \bibinfo{author}{\bibfnamefont{A.}~\bibnamefont{{Agnello}}}, \bibnamefont{and} 
  \bibinfo{author}{\bibfnamefont{N.~W.} \bibnamefont{{Evans}}}, 
  \bibinfo{journal}{Mon. Not. R. Astron. Soc.} \textbf{\bibinfo{volume}{429}}, 
  \bibinfo{pages}{L89} (\bibinfo{year}{2013}).

\bibitem[{\citenamefont{{Navarro} et~al.}(1997)\citenamefont{{Navarro}, {Frenk}, and {White}}}]{NFW1997}
  \bibinfo{author}{\bibfnamefont{J.~F.} \bibnamefont{{Navarro}}},
  \bibinfo{author}{\bibfnamefont{C.~S.} \bibnamefont{{Frenk}}}, \bibnamefont{and}
  \bibinfo{author}{\bibfnamefont{S.~D.~M.} \bibnamefont{{White}}},
  \bibinfo{journal}{Astrophys. J.} \textbf{\bibinfo{volume}{490}},
  \bibinfo{pages}{493} (\bibinfo{year}{1997}).

\bibitem[{\citenamefont{{Bode} et~al.}(2001)\citenamefont{{Bode}, {Ostriker}, and {Turok}}}]{Bode2001}
  \bibinfo{author}{\bibfnamefont{P.}~\bibnamefont{{Bode}}},
  \bibinfo{author}{\bibfnamefont{J.~P.} \bibnamefont{{Ostriker}}}, \bibnamefont{and}
  \bibinfo{author}{\bibfnamefont{N.}~\bibnamefont{{Turok}}},
  \bibinfo{journal}{Astrophys. J.} \textbf{\bibinfo{volume}{556}},
  \bibinfo{pages}{93} (\bibinfo{year}{2001}).

\bibitem[{\citenamefont{{Klypin} et~al.}(1999)\citenamefont{{Klypin},
  {Kravtsov}, {Valenzuela}, and {Prada}}}]{Klypin1999}
\bibinfo{author}{\bibfnamefont{A.}~\bibnamefont{{Klypin}}},
  \bibinfo{author}{\bibfnamefont{A.~V.} \bibnamefont{{Kravtsov}}},
  \bibinfo{author}{\bibfnamefont{O.}~\bibnamefont{{Valenzuela}}},
  \bibnamefont{and} \bibinfo{author}{\bibfnamefont{F.}~\bibnamefont{{Prada}}},
  \bibinfo{journal}{Astrophys. J.} \textbf{\bibinfo{volume}{522}},
  \bibinfo{pages}{82} (\bibinfo{year}{1999}).

\bibitem[{\citenamefont{{Moore} et~al.}(1999)\citenamefont{{Moore}, {Ghigna},
  {Governato}, {Lake}, {Quinn}, {Stadel}, and {Tozzi}}}]{Moore1999}
\bibinfo{author}{\bibfnamefont{B.}~\bibnamefont{{Moore}}},
  \bibinfo{author}{\bibfnamefont{S.}~\bibnamefont{{Ghigna}}},
  \bibinfo{author}{\bibfnamefont{F.}~\bibnamefont{{Governato}}},
  \bibinfo{author}{\bibfnamefont{G.}~\bibnamefont{{Lake}}},
  \bibinfo{author}{\bibfnamefont{T.}~\bibnamefont{{Quinn}}},
  \bibinfo{author}{\bibfnamefont{J.}~\bibnamefont{{Stadel}}}, \bibnamefont{and}
  \bibinfo{author}{\bibfnamefont{P.}~\bibnamefont{{Tozzi}}},
  \bibinfo{journal}{Astrophys. J.} \textbf{\bibinfo{volume}{524}},
  \bibinfo{pages}{L19} (\bibinfo{year}{1999}).

\bibitem[{\citenamefont{{Boylan-Kolchin} et~al.}(2011)\citenamefont{{Boylan-Kolchin}, {Bullock}, and {Kaplinghat}}}]{Boylan-KolchinBullockKaplinghat2011}
  \bibinfo{author}{\bibfnamefont{M.} \bibnamefont{{Boylan-Kolchin}}},
  \bibinfo{author}{\bibfnamefont{J.~S.} \bibnamefont{{Bullock}}}, \bibnamefont{and}
  \bibinfo{author}{\bibfnamefont{M.} \bibnamefont{{Kaplinghat}}},
  \bibinfo{journal}{Mon. Not. R. Astron. Soc.} \textbf{\bibinfo{volume}{415}},
  \bibinfo{pages}{L40} (\bibinfo{year}{2011}).

\bibitem[{\citenamefont{{Papastergis} et~al.}(2015)\citenamefont{{Papastergis}, {Giovanelli}, {Haynes}, and {Shankar}}}]{Papastergis2015}
  \bibinfo{author}{\bibfnamefont{E.} \bibnamefont{{Papastergis}}},
  \bibinfo{author}{\bibfnamefont{R.} \bibnamefont{{Giovanelli}}},
  \bibinfo{author}{\bibfnamefont{M.~P.} \bibnamefont{{Haynes}}}, \bibnamefont{and}
  \bibinfo{author}{\bibfnamefont{F.} \bibnamefont{{Shankar}}},
 \bibinfo{journal}{A\&A} \textbf{\bibinfo{volume}{574}},
  \bibinfo{pages}{A113} (\bibinfo{year}{2015}).

\bibitem[{\citenamefont{{Marsh}, and {Silk}}(2014)}]{MarshSilk2014}
  \bibinfo{author}{\bibfnamefont{D.~J.E.}~\bibnamefont{{Marsh}}}, \bibnamefont{and}
  \bibinfo{author}{\bibfnamefont{J.}~\bibnamefont{{Silk}}},
  \bibinfo{journal}{Mon. Not. R. Astron. Soc.} \textbf{\bibinfo{volume}{437}},
  \bibinfo{pages}{2652} (\bibinfo{year}{2014}).

\bibitem[{\citenamefont{Schive et~al.}(2016)\citenamefont{{Schive}, {Chiueh}, {Broadhurst} and {Huang}}}]{Schive2016}
  \bibinfo{author}{\bibfnamefont{H.-Y.} \bibnamefont{Schive}},
  \bibinfo{author}{\bibfnamefont{T.}~\bibnamefont{Chiueh}},
  \bibinfo{author}{\bibfnamefont{T.}~\bibnamefont{Broadhurst}}, \bibnamefont{and}
  \bibinfo{author}{\bibfnamefont{K.-W.} \bibnamefont{Huang}},
  \bibinfo{journal}{Astrophys. J.} \textbf{\bibinfo{volume}{818}},
  \bibinfo{pages}{89} (\bibinfo{year}{2016}).

\bibitem[{\citenamefont{{Macci{\`o}} et~al.}(2012)\citenamefont{{Macci{\`o}}, {Paduroiu}, {Anderhalden}, {Schneider}, and {Moore}}}]{Maccio2012}
  \bibinfo{author}{\bibfnamefont{A.~V.} \bibnamefont{{Macci{\`o}}}},
  \bibinfo{author}{\bibfnamefont{S.}~\bibnamefont{{Paduroiu}}},
  \bibinfo{author}{\bibfnamefont{D.}~\bibnamefont{{Anderhalden}}},
  \bibinfo{author}{\bibfnamefont{A.}~\bibnamefont{{Schneider}}}, \bibnamefont{and}
  \bibinfo{author}{\bibfnamefont{B.}~\bibnamefont{{Moore}}},
  \bibinfo{journal}{Mon. Not. R. Astron. Soc.} \textbf{\bibinfo{volume}{424}},
  \bibinfo{pages}{1105} (\bibinfo{year}{2012}).

\bibitem[{\citenamefont{{Woo}, and {Chiueh}}(2009)}]{WooChiueh2009}
  \bibinfo{author}{\bibfnamefont{T.~P.}~\bibnamefont{{Woo}}}, \bibnamefont{and}
  \bibinfo{author}{\bibfnamefont{T.}~\bibnamefont{{Chiueh}}},
  \bibinfo{journal}{Astrophys. J.} \textbf{\bibinfo{volume}{697}},
  \bibinfo{pages}{850} (\bibinfo{year}{2009}).

\bibitem[{\citenamefont{{Hlozek} et~al.}(2015)\citenamefont{{Hlozek}, {Grin}, {Marsh}, and {Ferreira}}}]{Hlozek2015}
  \bibinfo{author}{\bibfnamefont{R.}~\bibnamefont{{Hlozek}}},
  \bibinfo{author}{\bibfnamefont{D.}~\bibnamefont{{Grin}}},
  \bibinfo{author}{\bibfnamefont{D.~J.E.}~\bibnamefont{{Marsh}}}, \bibnamefont{and}
  \bibinfo{author}{\bibfnamefont{P.~G.}~\bibnamefont{{Ferreira}}},
  \bibinfo{journal}{Phys. Rev. D.} \textbf{\bibinfo{volume}{91}},
  \bibinfo{pages}{103512} (\bibinfo{year}{2015}).

\bibitem[{\citenamefont{{Marsh}}(2016)}]{Marsh2016}
  \bibinfo{author}{\bibfnamefont{D.~J.E.}~\bibnamefont{{Marsh}}},
  \bibinfo{journal}{Phys. Rep.} \textbf{\bibinfo{volume}{643}},
  \bibinfo{pages}{1} (\bibinfo{year}{2016}).

\bibitem[{\citenamefont{{Chiueh}}(2014)}]{ChiueharXiv2014}
  \bibinfo{author}{\bibfnamefont{T.}~\bibnamefont{{Chiueh}}},
  {arXiv:1409.0380} (\bibinfo{year}{2014}).

\bibitem[{\citenamefont{{Ma}, and {Bertschinger}}(1995)}]{MaBertschinger1995}
  \bibinfo{author}{\bibfnamefont{C.-P.}~\bibnamefont{{Ma}}}, \bibnamefont{and}
  \bibinfo{author}{\bibfnamefont{E.}~\bibnamefont{{Bertschinger}}},
  \bibinfo{journal}{Astrophys. J.} \textbf{\bibinfo{volume}{455}},
  \bibinfo{pages}{7} (\bibinfo{year}{1995}).

\bibitem[{\citenamefont{{Planck Collaboration}
  et~al.}(2016)\citenamefont{{Planck Collaboration}, {Ade}, {Aghanim}, {Arnaud},
  {Ashdown}, {Aumont}, {Baccigalupi}, {Banday}, {Barreiro}, {Bartlett}
  et~al.}}]{Planck2016}
\bibinfo{author}{\bibnamefont{{Planck Collaboration}}},
  \bibinfo{author}{\bibfnamefont{P.~A.~R.} \bibnamefont{{Ade}}},
  \bibinfo{author}{\bibfnamefont{N.}~\bibnamefont{{Aghanim}}},
  \bibinfo{author}{\bibfnamefont{M.}~\bibnamefont{{Arnaud}}},
  \bibinfo{author}{\bibfnamefont{M.}~\bibnamefont{{Ashdown}}},
  \bibinfo{author}{\bibfnamefont{J.}~\bibnamefont{{Aumont}}},
  \bibinfo{author}{\bibfnamefont{C.}~\bibnamefont{{Baccigalupi}}},
  \bibinfo{author}{\bibfnamefont{A.~J.} \bibnamefont{{Banday}}},
  \bibinfo{author}{\bibfnamefont{R.~B.} \bibnamefont{{Barreiro}}},
  \bibinfo{author}{\bibfnamefont{J.~G.} \bibnamefont{{Bartlett}}},
  \bibnamefont{et~al.}, \bibinfo{journal}{A\&A} \textbf{\bibinfo{volume}{594}},
  \bibinfo{pages}{A13} (\bibinfo{year}{2016}).

\bibitem[{\citenamefont{{Hu}, and {Sugiyama}}(1996)}]{HuSugiyama1996}
  \bibinfo{author}{\bibfnamefont{W.}~\bibnamefont{{Hu}}}, \bibnamefont{and}
  \bibinfo{author}{\bibfnamefont{N.}~\bibnamefont{{Sugiyama}}},
  \bibinfo{journal}{Astrophys. J.} \textbf{\bibinfo{volume}{471}},
  \bibinfo{pages}{542} (\bibinfo{year}{1996}).

\bibitem[{\citenamefont{{Zhang} and {Chiueh}}(2017)}]{ZhangChiueharXiv2017}
  \bibinfo{author}{\bibfnamefont{U.-H.} \bibnamefont{{Zhang}}} \bibnamefont{and}
  \bibinfo{author}{\bibfnamefont{T.}~\bibnamefont{{Chiueh}}},
  {arXiv:1705.01439} (\bibinfo{year}{2017}).

\bibitem[{\citenamefont{{Bouwens} et~al.}(2011)}]{Bouwens2011}
  \bibinfo{author}{\bibfnamefont{R.~J.}~\bibnamefont{{Bouwens}}},
  \bibinfo{author}{\bibfnamefont{G.~D.}~\bibnamefont{{Illingworth}}},
  \bibinfo{author}{\bibfnamefont{I.}~\bibnamefont{{Labbe}}},
  \bibinfo{author}{\bibfnamefont{P.~A.}~\bibnamefont{{Oesch}}},
  \bibinfo{author}{\bibfnamefont{M.}~\bibnamefont{{Trenti}}},
  \bibinfo{author}{\bibfnamefont{C.~M.}~\bibnamefont{{Carollo}}}, \bibnamefont{et~al.},
  \bibinfo{journal}{Nature} \textbf{\bibinfo{volume}{469}},
  \bibinfo{pages}{504} (\bibinfo{year}{2011}).

\bibitem[{\citenamefont{{Oesch} et~al.}(2012)}]{Bouwens2012}
  \bibinfo{author}{\bibfnamefont{P.~A.}~\bibnamefont{{Oesch}}},
  \bibinfo{author}{\bibfnamefont{R.~J.}~\bibnamefont{{Bouwens}}},
  \bibinfo{author}{\bibfnamefont{G.~D.}~\bibnamefont{{Illingworth}}},
  \bibinfo{author}{\bibfnamefont{I.}~\bibnamefont{{Labb{\'e}}}},
  \bibinfo{author}{\bibfnamefont{M.}~\bibnamefont{{Trenti}}},
  \bibinfo{author}{\bibfnamefont{V.}~\bibnamefont{{Gonzalez}}},
  \bibinfo{author}{\bibfnamefont{C.~M.}~\bibnamefont{{Carollo}}},
  \bibinfo{author}{\bibfnamefont{M.}~\bibnamefont{{Franx}}},
  \bibinfo{author}{\bibfnamefont{P.~G.}~\bibnamefont{{van Dokkum}}}, \bibnamefont{and}
  \bibinfo{author}{\bibfnamefont{D.}~\bibnamefont{{Magee}}},
  \bibinfo{journal}{Astrophys. J.} \textbf{\bibinfo{volume}{745}},
  \bibinfo{pages}{110} (\bibinfo{year}{2012}).


\bibitem[{\citenamefont{{Zheng} et~al.}(2012)}]{Zheng2012}
  \bibinfo{author}{\bibfnamefont{W.}~\bibnamefont{{Zheng}}},
  \bibinfo{author}{\bibfnamefont{M.}~\bibnamefont{{Postman}}},
  \bibinfo{author}{\bibfnamefont{A.}~\bibnamefont{{Zitrin}}},
  \bibinfo{author}{\bibfnamefont{J.}~\bibnamefont{{Moustakas}}},
  \bibinfo{author}{\bibfnamefont{X.}~\bibnamefont{{Shu}}},
  \bibinfo{author}{\bibfnamefont{S.}~\bibnamefont{{Jouvel}}}, \bibnamefont{et~al.},
  \bibinfo{journal}{Nature} \textbf{\bibinfo{volume}{489}},
  \bibinfo{pages}{406} (\bibinfo{year}{2012}).

\bibitem[{\citenamefont{{Coe} et~al.}(2013)}]{Coe2013}
  \bibinfo{author}{\bibfnamefont{D.}~\bibnamefont{{Coe}}},
  \bibinfo{author}{\bibfnamefont{A.}~\bibnamefont{{Zitrin}}},
  \bibinfo{author}{\bibfnamefont{M.}~\bibnamefont{{Carrasco}}},
  \bibinfo{author}{\bibfnamefont{X.}~\bibnamefont{{Shu}}},
  \bibinfo{author}{\bibfnamefont{W.}~\bibnamefont{{Zheng}}},
  \bibinfo{author}{\bibfnamefont{M.}~\bibnamefont{{Postman}}}, \bibnamefont{et~al.},
  \bibinfo{journal}{Astrophys. J.} \textbf{\bibinfo{volume}{762}},
  \bibinfo{pages}{32} (\bibinfo{year}{2013}).

\bibitem[{\citenamefont{{Oesch} et~al.}(2014)}]{Oesch2014}
  \bibinfo{author}{\bibfnamefont{P.~A.}~\bibnamefont{{Oesch}}},
  \bibinfo{author}{\bibfnamefont{R.~J.}~\bibnamefont{{Bouwens}}},
  \bibinfo{author}{\bibfnamefont{G.~D.}~\bibnamefont{{Illingworth}}},
  \bibinfo{author}{\bibfnamefont{I.}~\bibnamefont{{Labb{\'e}}}},
  \bibinfo{author}{\bibfnamefont{R.}~\bibnamefont{{Smit}}},
  \bibinfo{author}{\bibfnamefont{M.}~\bibnamefont{{Franx}}}, \bibnamefont{et~al.},
  \bibinfo{journal}{Astrophys. J.} \textbf{\bibinfo{volume}{786}},
  \bibinfo{pages}{108} (\bibinfo{year}{2014}).

\bibitem[{\citenamefont{{Zitrin} et~al.}(2014)}]{Zitrin2014}
  \bibinfo{author}{\bibfnamefont{A.}~\bibnamefont{{Zitrin}}},
  \bibinfo{author}{\bibfnamefont{W.}~\bibnamefont{{Zheng}}},
  \bibinfo{author}{\bibfnamefont{T.}~\bibnamefont{{Broadhurst}}},
  \bibinfo{author}{\bibfnamefont{J.}~\bibnamefont{{Moustakas}}},
  \bibinfo{author}{\bibfnamefont{D.}~\bibnamefont{{Lam}}},
  \bibinfo{author}{\bibfnamefont{X.}~\bibnamefont{{Shu}}}, \bibnamefont{et~al.},
  \bibinfo{journal}{Astrophys. J. Lett.} \textbf{\bibinfo{volume}{793}},
  \bibinfo{pages}{L12} (\bibinfo{year}{2014}).

\bibitem[{\citenamefont{{Bouwens} et~al.}(2015)}]{Bouwens2015}
  \bibinfo{author}{\bibfnamefont{R.~J.}~\bibnamefont{{Bouwens}}},
  \bibinfo{author}{\bibfnamefont{G.~D.}~\bibnamefont{{Illingworth}}},
  \bibinfo{author}{\bibfnamefont{P.~A.}~\bibnamefont{{Oesch}}},
  \bibinfo{author}{\bibfnamefont{M.}~\bibnamefont{{Trenti}}},
  \bibinfo{author}{\bibfnamefont{I.}~\bibnamefont{{Labb{\'e}}}},
  \bibinfo{author}{\bibfnamefont{L.}~\bibnamefont{{Bradley}}}, \bibnamefont{et~al.},
  \bibinfo{journal}{Astrophys. J.} \textbf{\bibinfo{volume}{803}},
  \bibinfo{pages}{34} (\bibinfo{year}{2015}).

\bibitem[{\citenamefont{{Bernard} et~al.}(2016)}]{Bernard2016}
  \bibinfo{author}{\bibfnamefont{S.~R.}~\bibnamefont{{Bernard}}},
  \bibinfo{author}{\bibfnamefont{D.}~\bibnamefont{{Carrasco}}},
  \bibinfo{author}{\bibfnamefont{M.}~\bibnamefont{{Trenti}}},
  \bibinfo{author}{\bibfnamefont{P.~A.}~\bibnamefont{{Oesch}}},
  \bibinfo{author}{\bibfnamefont{J.~F.}~\bibnamefont{{Wu}}},
  \bibinfo{author}{\bibfnamefont{L.~D.}~\bibnamefont{{Bradley}}}, \bibnamefont{et~al.},
  \bibinfo{journal}{Astrophys. J.} \textbf{\bibinfo{volume}{827}},
  \bibinfo{pages}{76} (\bibinfo{year}{2016}).

\end{thebibliography}

\renewcommand{\theequation}{A\arabic{equation}}
\setcounter{equation}{0}
\appendix
\renewcommand{\theequation}{A\arabic{equation}}
\section{Passive Evolution}
\label{app: passive evolution}

In this appendix, governing equations and analytic solutions for passive evolution are presented.

The action can be written as follows,
\begin{equation}
\label{equ: action}
S= \int d^4x\sqrt{|g|} \Big [ {R\over{16 \pi G}} + L_M \Big ],
\end{equation}
where $g$ is the determinant of the metric $g_{\mu\nu}$, $R$ the scalar curvature, $L_M$ the Lagrangian density of matter and $G$ the gravitational constant. Here we choose the speed of light and the Planck constant $\hbar$ to be $1$.

The Lagrangian density $L_M$ consists of two components, the radiation fluid and the dark matter. The radiation can be described as a perfect fluid with the equation of state $P=\epsilon/3$, where $P$ is the pressure and $\epsilon$ is the energy density. On the other hand, two different dark matter models, the cold dark matter (CDM) model and the $\psi$ dark matter ($\psi$DM), will be introduced. CDM is a pressureless perfect fluid, and $\psi$DM is a spin-zero real scalar field with a Lagrangian density
\begin{equation}
\label{equ: lagrangain density of psi}
L_{\psi} = {1\over2}g^{\mu\nu}\partial_{\mu}\psi\partial_{\nu}\psi - V(\psi),
\end{equation}
where $V$ is the scalar field potential. The corresponding stress-energy tensor is
\begin{equation}
\label{equ: stress-energy tensor for psi}
T_{\mu\nu}^{\psi} = {{-2}\over{\sqrt{|g|}}}{{\delta(\sqrt{|g|}L_{\psi})}\over{\delta g^{\mu\nu}}}.
\end{equation}

The metric is the flat Friedmann-Lemaitre-Robertson-Walker (FLRW) metric with the scalar metric perturbation, i.e.,
\begin{equation}
\label{equ: metric}
ds^2= a^2(\tau) \{ [1+2\phi(x^i,\tau)]d\tau^2-[1-2\phi(x^i,\tau)]\delta_{ij}dx^idx^j \},
\end{equation}
where $\tau$ is the conformal time, $x^i$ with $i=1,2,3$ is the comoving coordinate, $\phi$ the metric perturbation, $a$ the scale factor, $\delta_{ij}$ the Kronecher delta function, and the Newtonian gauge has been chosen. Similarly, the energy-momentum stress tensor is also written as a uniform background with a small perturbation. Thus $\psi$DM field can be expressed as $\Psi(\tau) + \psi(x^i,\tau)$ for calculating the corresponding stress-energy tensor.

Gathering the above and substituting into Eq. (\ref{equ: action}) and applying Euler-Lagrange equation, one obtains the Einstein equations and the conservation of energy and momentum.

The zeroth order equations are just Friedmann equations,
\begin{equation}
\label{equ: Friedmann equations 1}
H^2 = {{8\pi G}\over{3}} a^2\Big ( \epsilon_\gamma + \epsilon_D \Big ),
\end{equation}
\begin{equation}
\label{equ: Friedmann equations 2}
H^{'}-H^2 = -4\pi G a^2 [ ( \epsilon_\gamma + P_\gamma ) + ( \epsilon_D + P_D ) ],
\end{equation}
with the following conservation laws,
\begin{equation}
\label{equ: conservation}
\epsilon_\alpha^{'}+3H ( \epsilon_\alpha + P_\alpha ) = 0, \text{  } \alpha=\gamma,\text{ } D,
\end{equation}
where $()^{'}$ is the derivative with respect to $\tau$, $H \equiv a^{'}/a$ the Hubble parameter, and the subscript denotes different components. Here we use $\gamma$ to denote the radiation fluid and $D$ the dark matter.

For the CDM model, the pressure $P_D$ is zero, and Eq. (\ref{equ: conservation}) becomes,
\begin{equation}
\label{equ: zeroth order equation for CDM}
\epsilon_C^{'}+3H\epsilon_C = 0,
\end{equation}
where we have changed the subscript "$D$" to "$C$" to label CDM. For the $\psi$DM model, we have the following relations
\begin{equation}
\label{equ: zeroth order energy density for psiDM}
\epsilon_{\psi} = {1\over2}{{(\Psi^{'})^2}\over{a^2}} + V(\Psi),
\end{equation}
\begin{equation}
\label{equ: zeroth order pressure for psiDM}
P_{\psi} = {1\over2}{{(\Psi^{'})^2}\over{a^2}} - V(\Psi),
\end{equation}
where the subscript "$\psi$" is used to replace "$D$". These relations can be obtained by Eq. (\ref{equ: stress-energy tensor for psi}). Substituting Eqs. (\ref{equ: zeroth order energy density for psiDM}) and (\ref{equ: zeroth order pressure for psiDM}) into Eq. (\ref{equ: conservation}), it follows that
\begin{equation}
\label{equ: zeroth order equation for psiDM}
\Psi^{''} + 2H\Psi^{'} + a^2{{dV}\over{d\psi}} \Big |_{\Psi} = 0.
\end{equation}

The first order perturbed equations consist of the perturbed Einstein equations
\begin{equation}
\label{equ: perturbed Einstein equation 1}
-k^2\phi -3H(\phi^{'}+H\phi) = 4\pi Ga^2(\delta\epsilon_{\gamma}+\delta\epsilon_{D}),
\end{equation}
\begin{equation}
\label{equ: perturbed Einstein equation 2}
\phi^{'}+H\phi = -4\pi Ga^2[(\epsilon_\gamma + P_\gamma)\theta_\gamma+(\epsilon_D + P_D)\theta_D],
\end{equation}
the continuity equation
\begin{equation}
\label{equ: continuity equation}
\left.\begin{aligned}
& \delta_{\alpha}^{'}+3H \Big ( {{\delta P_\alpha}\over{\delta\epsilon_\alpha}}-{{P_\alpha}\over{\epsilon_\alpha}} \Big )\delta_{\alpha} = \Big ( 1 + {{P_\alpha}\over{\epsilon_\alpha}} \Big )(k^2\theta_\alpha+3\phi^{'}), \\
& \text{  } \alpha = \gamma,\text{ } D,
\end{aligned}
\right.
\end{equation}
and the Euler equation
\begin{equation}
\label{equ: Euler equation}
\theta_{\alpha}^{'}+3H \Big ( {{1}\over{3}}-{{P_\alpha^{'}}\over{\epsilon_\alpha^{'}}} \Big )\theta_{\alpha} = - {{\delta P_\alpha}\over{\epsilon_\alpha+P_\alpha}} - \phi, \text{  } \alpha = \gamma,\text{ } D.
\end{equation}
Here $\delta\epsilon_\alpha$, $\delta P_\alpha$ and $\theta_\alpha$ are the energy density perturbation, the pressure perturbation and the velocity potential of the $\alpha$ species, respectively. The quantity $\delta_{\alpha}$ is the fractional overdensity of the $\alpha$ species defined as $\delta_\alpha \equiv \delta\epsilon_\alpha/\epsilon_\alpha$.  The spatial Fourier transformation has been applied in the comoving coordinate on Eqs. (\ref{equ: perturbed Einstein equation 1}), (\ref{equ: perturbed Einstein equation 2}), (\ref{equ: continuity equation}) and (\ref{equ: Euler equation}).

For the CDM model, $P_D = \delta P_D = 0$, Eqs. (\ref{equ: continuity equation}) and (\ref{equ: Euler equation}) become
\begin{equation}
\label{equ: continuity equation for CDM}
\delta_C^{'} = k^2\theta_C + 3\phi^{'},
\end{equation}
\begin{equation}
\label{equ: Euler equation for CDM}
\theta_C^{'} + H\theta_C = -\phi.
\end{equation}
On the other hand for $\psi$DM, from Eq. (\ref{equ: stress-energy tensor for psi}), the quantities $\delta\epsilon_D$, $\delta P_D$ and $\theta_D$ can be expressed as
\begin{equation}
\label{equ: energy density perturbation for psiDM}
\delta\epsilon_{\psi} = {{\Psi^{'}\psi^{'}}\over{a^2}} + {{dV}\over{d\psi}} \Big |_{\Psi}\psi - {{(\Psi^{'})^2\phi}\over{a^2}},
\end{equation}
\begin{equation}
\label{equ: pressure for psiDM}
\delta P_{\psi} = {{\Psi^{'}\psi^{'}}\over{a^2}} - {{dV}\over{d\psi}} \Big |_{\Psi}\psi - {{(\Psi^{'})^2\phi}\over{a^2}},
\end{equation}
\begin{equation}
\label{equ: velocity potential for psiDM}
\Psi^{'}\theta_{\psi} = -\psi.
\end{equation}
Substituting Eqs. (\ref{equ: energy density perturbation for psiDM}), (\ref{equ: pressure for psiDM}) and (\ref{equ: velocity potential for psiDM}) into Eq. (\ref{equ: continuity equation}), it follows that,
\begin{equation}
\label{equ: perturbed field equation for psiDM}
\psi^{''} + 2H\psi^{'} + \Big ( k^2 + a^2{{d^2 V}\over{d\psi}^2} \Big |_{\Psi} \Big )\psi = 4\Psi^{'}\phi^{'} -2a^2{{dV}\over{d\psi}} \Big |_{\Psi}\phi.
\end{equation}
In addition, the Euler equation (Eq. (\ref{equ: Euler equation})) is reduced to Eq. (\ref{equ: zeroth order equation for psiDM}), as shown in Sec. (\ref{sec: Governing Equations}).

In the radiation-dominant era, the dark matter energy density is much smaller than the radiation energy density, so are the perturbed quantities. Under this condition, dark matter can be treated as passive particles (fields) tracing the geometry of universe that is determined by radiation, and therefore the dark matter contribution in Friedmann equations (Eq. (\ref{equ: Friedmann equations 1}) and (\ref{equ: Friedmann equations 2})) and perturbed Einstein equations (Eqs. (\ref{equ: perturbed Einstein equation 1}) and (\ref{equ: perturbed Einstein equation 2})) can be ignored.

The form of scalar field potential $V(\psi)$ should be specified. For a non-self-interacting free field, we have $V(\psi)=m^2\psi^2/2$, where $m$ is the mass of the dark matter particle.

First, we should solve the zeroth order equations. From Eq. (\ref{equ: conservation}) for the photon fluid, the photon energy density $\epsilon_\gamma \propto a^{-4}$ due to $3P_\gamma=\epsilon_\gamma$. Substituting the above relation into Friedmann's equations, we find $a \propto \tau$ and $H \propto a^{-1}$.

When the CDM model is considered, it is straightforward to find the energy density $\epsilon_C \propto a^{-3}$. For the $\psi$DM model, substituting above scalar potential $V(\psi)$ into Eq. (\ref{equ: zeroth order equation for psiDM}), Eq. (\ref{equ: zeroth-order dark matter}) follows. It is natural to define a dimensionless quantity $u_m=ma/2H$ which normalizes the time to the onset of mass oscillation. With above relations, Eq. (\ref{equ: zeroth-order dark matter}) becomes
\begin{equation}
\label{equ: zeroth-order dark matter 1}
{{d^2\Psi}\over{du_m^2}}+{3\over{2u_m}}{{d\Psi}\over{du_m}} + \Psi=0.
\end{equation}
Equation (\ref{equ: zeroth-order dark matter 1}) has following solution,
\begin{equation}
\label{equ: solution for zeroth-order dark matter}
\Psi(u_m)=C^{\Psi}_r\Psi_r(u_m) + C^{\Psi}_s\Psi_s(u_m),
\end{equation}
where $C^{\Psi}_r$ and $C^{\Psi}_s$ are constants, and $\Psi_r$ and $\Psi_s$ are
\begin{equation}
\label{equ: Psi_r}
\Psi_r(u_m)=2^{{1\over4}}\Gamma \Big ( {5\over4} \Big ) {{J_{1\over4}({u_m})}\over{u_m^{1\over4}}},
\end{equation}
\begin{equation}
\label{equ: Psi_s}
\Psi_s(u_m)=2^{{1\over4}}\Gamma \Big ( {5\over4} \Big ){{Y_{1\over4}({u_m})}\over{u_m^{1\over4}}},
\end{equation}
where $\Gamma$ is the gamma function, $J$ and $Y$ are Bessel function of the first and second kind, respectively, which are orthogonal to each other. It is noted that $\Psi_s$ has a singularity as $u_m\rightarrow 0$ while $\Psi_r$ remains regular, and hence only $\Psi_r$ is the valid zero-order field.

When $u_m\gg 1$, the energy density $\propto u_m^{-3/2} \propto a^{-3}$. This means the energy density of $\psi$DM follows CDM after the mass oscillation.
We further define $\Psi_0 \equiv C^{\Psi}_r\Psi_r(0)=C^{\Psi}_r$. In the following, $C^{\Psi}_r$ will be replaced by $\Psi_0$.

Next, we turn to the first order perturbed equations. The metric perturbation $\phi$ should be obtained first. To solve $\phi$, it is natural to define a dimensionless quantity $u_k \equiv k/\sqrt{3}H$. This quantity approximately measures the epoch of the horizon entry. Using the chain rule, Eq. (\ref{equ: perturbed Einstein equation 2}) becomes
\begin{equation}
\label{equ: metric perturbation 3}
{{d\phi}\over{du_k}} + {{\phi}\over{u_k}} +  {{g_\gamma}\over{2u_k}} = 0,
\end{equation}
where $g_\gamma \equiv 4H\theta_\gamma$. On the other hand, from Eqs. (\ref{equ: Friedmann equations 1}), (\ref{equ: Friedmann equations 2}), (\ref{equ: perturbed Einstein equation 1}), (\ref{equ: perturbed Einstein equation 2}), (\ref{equ: continuity equation}) and (\ref{equ: Euler equation}) with the radiation fluid's equation of state $P_\gamma = \epsilon_\gamma/3$ and $\delta P_\gamma = \delta \epsilon_\gamma/3$, $g_\gamma$ obeys the following equation
\begin{equation}
\label{equ: g_gamma equation}
{{dg_\gamma}\over{du_k}} = \Big ( 2u_k - {4\over{u_k}} \Big )\phi -{2\over{u_k}}g_\gamma.
\end{equation}
Combining Eq. (\ref{equ: metric perturbation 3}) and (\ref{equ: g_gamma equation}), one finds $\phi$ satisfies the following second order equation
\begin{equation}
\label{equ: metric equation}
{{d^2\phi}\over{du_k^2}} + {4\over u_k}{{d\phi}\over{du_k}} + \phi = 0.
\end{equation}
Equation (\ref{equ: metric equation}) has a solution:
\begin{equation}
\label{equ: metric equation solution}
\phi = C^{\phi}_r\phi_r(u_k) + C^{\phi}_s\phi_s(u_k),
\end{equation}
where $C^{\phi}_r$ and $C^{\phi}_s$ are constants, and
\begin{equation}
\label{equ: phi_r}
\phi_r(u_k) = 3 \Big [ -{{\cos(u_k)}\over{u_k^2}} + {{\sin(u_k)}\over{u_k^3}} \Big ],
\end{equation}
\begin{equation}
\label{equ: phi_s}
\phi_s(u_k) = 3 \Big [ -{{\sin(u_k)}\over{u_k^2}} - {{\cos(u_k)}\over{u_k^3}} \Big ].
\end{equation}

From Eqs. (\ref{equ: continuity equation}) and (\ref{equ: Euler equation}), the radiation energy perturbation $\delta_\gamma$ becomes,
\begin{equation}
\label{equ: photon energy perturbation solution}
\delta_\gamma = C^{\phi}_r\delta_\gamma^r(u_k) + C^{\phi}_s\delta_\gamma^s(u_k),
\end{equation}
with
\begin{equation}
\label{equ: delta_gamma^r}
\delta_\gamma^r(u_k)=3 \Big [ 2\cos(u_k) - 4{{\sin(u_k)}\over{u_k}} - 4{{\cos(u_k)}\over{u_k^2}} + 4{{\sin(u_k)}\over{u_k^3}} \Big ],
\end{equation}
\begin{equation}
\label{equ: delta_gamma^s}
\delta_\gamma^s(u_k)=3 \Big [ 2\sin(u_k) + 4{{\cos(u_k)}\over{u_k}} - 4{{\sin(u_k)}\over{u_k^2}} - 4{{\cos(u_k)}\over{u_k^3}} \Big ].
\end{equation}

Since $\phi_s$ and $\delta_\gamma^s$ have a singularity at $u_k=0$, only the regular modes, $\phi_r$ and $\delta_\gamma^r$, are retained. Similar to the $\psi$DM background field case, we can define $\phi_0 \equiv C^{\phi}_r\phi_r(0) = C^{\phi}_r$. In the following, $C^{\phi}_r$ will be replaced by $\phi_0$.

It is noted that $\phi(u_k)= \phi_0(1 + O(u_k^2))$, dominated by a constant $\phi_0$. On the other hand, from Eq. (\ref{equ: Poisson equation}) ignoring the dark matter contribution, the gauge covariant photon energy density perturbation grows as $\Delta_\gamma \propto u_k^2 \propto a^2$ before entering horizon.

Moreover, from Eq. (\ref{equ: delta_gamma^r}), $\delta_\gamma(u_k) = \phi_0 ( -2 + O(u_k^2) )$, when $u_k \ll 1$. This relation is used in the discussion of phase (\rmnum{1}a) in Sec. (\ref{sec: Passive Evolution and Asymptotic Solutions}).

The perturbed dark matter equations are solvable once the background solution and the metric perturbation are given. For CDM model, Eqs. (\ref{equ: continuity equation for CDM}) and (\ref{equ: Euler equation for CDM}) become
\begin{equation}
\label{equ: CDM fractional overdensity equation}
{{d\delta_C}\over{du_k}} = u_k g_C +  3{{d\phi}\over{du_k}},
\end{equation}
\begin{equation}
\label{equ: CDM velocity potential equation}
{{dg_C}\over{du_k}} + 2{{g_C}\over{u_k}} =  -3{{\phi}\over{u_k}},
\end{equation}
where $g_C \equiv 3H\theta_C$ and the metric perturbation $\phi= \phi_0 \phi_r(u_k)$. It is straightforward to solve Eq. (\ref{equ: CDM velocity potential equation}) for the solution of $g_C$, as
\begin{equation}
\label{equ: CDM velocity potential solution}
g_C(u_k)=-9\phi_0 \Big [ {1\over{u_k^2}} - {{\sin(u_k)}\over{u_k^3}}\Big ].
\end{equation}
Here the homogeneous solution of $g_C$ has been dropped since it has the singularity at $u_k=0$. Substituting Eq. (\ref{equ: CDM velocity potential solution}) into Eq. (\ref{equ: CDM fractional overdensity equation}) and integrating on both sides, one obtains
\begin{equation}
\label{equ: CDM fractional overdensity solution}
\left.\begin{aligned}
\delta_C(u_k)=-9\phi_0\Big [ & {{\sin(u_k)}\over{u_k}} + {{\cos(u_k)}\over{u_k^2}} - {{\sin(u_k)}\over{u_k^3}} + \\
                              & \ln(u_k) - Ci(u_k) + \gamma - {1\over2}\Big ],
\end{aligned}
\right.
\end{equation}
where $Ci(u_k)$ is the cosine integral function and $\gamma$ is the Euler-Mascheroni constant. The last term, $1/2$, is fixed by the adiabatic condition (Eq. (\ref{equ: adiabatic condition})). Finally, the gauge covariant energy perturbation $\Delta_C$ is obtained to be
\begin{equation}
\label{equ: CDM gauge covariant energy density solution}
\left.\begin{aligned}
\Delta_C = \delta_C - g_C = -9\phi_0 \Big [ & {{\sin(u_k)}\over{u_k}} + {{\cos(u_k)-1}\over{u_k^2}} + \\
                                            & \ln(u_k) - Ci(u_k) + \gamma - {1\over2} \Big ].
\end{aligned}
\right.
\end{equation}

For $\psi$DM model, Eq. (\ref{equ: perturbed field equation for psiDM}) becomes Eq. (\ref{equ: first-order dark matter})for free particle. Using $u_m$ as the argument, Eq. (\ref{equ: first-order dark matter}) further becomes
\begin{equation}
\label{equ: perturbed dark matter equation}
\left.\begin{aligned}
{{d^2\psi}\over{du_m^2}} + {3\over{2u_m}}{{d\psi}\over{du_m}} + \Big ( 1+ {A_k\over u_m} \Big )\psi & = 2 \Big [ {{u_k}\over{u_m}}{{d\phi}\over{du_k}}{{d\Psi}\over{du_m}} - \phi\Psi \Big ] \\
                                                                                               & \equiv S(u_m),
\end{aligned}
\right.
\end{equation}
where $A_k = k^2/(2maH)$, a dimensionless constant, and $S$ is the driving source. We note that $u_k^2=4A_k u_m/3$. Equation (\ref{equ: perturbed dark matter equation}) has following solution,
\begin{equation}
\label{equ: solution of perturbed dark matter equation}
\psi = C^{\psi}_r \psi_r(u_m) + C^{\psi}_s \psi_s(u_m) + \psi_p(u_m).
\end{equation}
Here $C^{\psi}_r$ and $C^{\psi}_s$ are constants, $\psi_r$ and $\psi_s$ are the homogeneous solutions of Eq. (\ref{equ: perturbed dark matter equation}) with the following forms,
\begin{equation}
\label{equ:psi_r}
\psi_r(u_m) = e^{-iu_m}M({3\over4} + {A_k\over2}i, {3\over2}, i2u_m),
\end{equation}
\begin{equation}
\label{equ:psi_s}
\psi_s(u_m) = e^{-iu_m}{{M({1\over4} + {A_k\over2}i, {1\over2}, i2u_m)}\over{\sqrt{u_m}}},
\end{equation}
where $M$ is the Kummer's function, and
\begin{equation}
\label{equ:psi_p}
\psi_p(u_m) = \int_0^{u_m}{{\psi_r(x)\psi_s(u_m)-\psi_r(u_m)\psi_s(x)}\over{W(\psi_r,\psi_s)(x)}}S(x)dx,
\end{equation}
the particular integral of Eq.(\ref{equ: perturbed dark matter equation}) with initial conditions $\psi_p(0)=d\psi_p/du_m(0)=0$. Here $W(\psi_r,\psi_s)(u_m)=\psi_r d\psi_s/du_m - \psi_s d\psi_r/du_m = -u_m^{-3/2}/2$, the Wronskian of $\psi_r$ and $\psi_s$.

Again, the constants, $C^{\psi}_r$ and $C^{\psi}_s$, can be determined by the adiabatic condition for the super-horizon mode. Particularly, these two constants must be zero if only regular mode of metric perturbation is considered. Therefore $\psi = \psi_p$, representing the adiabatic perturbation.

Once the perturbed field's solution is obtained, the gauge covariant energy density perturbation follows from Eq. (\ref{equ: covariant energy}), and becomes
\begin{equation}
\label{equ: comoving gauge psiDM density perturbation}
\Delta_\psi = { { {{d\Psi}\over{du_m}}{{d\psi_p}\over{du_m}} + \Psi\psi_p - \Big ({{d\Psi}\over{du_m}} \Big )^2\phi + {3\over2}{{{{d\Psi}\over{du_m}}\psi_p}\over{u_m}} }\over{ {1\over2} \Big [ \Big ({{d\Psi}\over{du_m}} \Big )^2 + \Psi^2 \Big ] }}.
\end{equation}

In the following, the asymptotic behavior of the gauge covariant dark matter energy density perturbation is to be derived. The CDM model is considered first. From Eq. (\ref{equ: CDM gauge covariant energy density solution}), it is straightforward to find
\begin{equation}
\label{equ: CDM gauge covariant energy density asymptotic form}
\Delta_C = -9\phi_0
\begin{cases}
{{u_k^2}\over8} + O(u_k^4), &\text{if } u_k \ll 1 \\
\ln(u_k) + O(1),            &\text{if } u_k \gg 1.
\end{cases}
\end{equation}

For $\psi$DM model, it is more complicated and has four different phases as discussed in Sec. (\ref{sec: Passive Evolution and Asymptotic Solutions}), which are to be verified from Eqs. (\ref{equ: solution of perturbed dark matter equation}) to (\ref{equ:psi_p}). In the following presentation, we will organize the four phases in the chronological order. The phases before mass oscillation (phase (\rmnum{2}) in Sec. (\ref{sec: Passive Evolution and Asymptotic Solutions})) are dealt with first and then we will show the results of the phases after mass oscillation (phase (\rmnum{1})).

\bigskip

(A) Before mass oscillation and super-horizon mode (phase (\rmnum{2}a))

\medskip

This phase can be dealt with using Eq. (\ref{equ:psi_p}) directly. The background field $\Psi$ has the following asymptotic form $\Psi = \Psi_0[1 - {u_m^2/5} + O(u_m^4)]$, and the metric perturbation $\phi$ can be expressed $\phi = \phi_0[1 - u_k^2/10 + O(u_k^4)] = \phi_0[1 - 2A_k u_m/15 + O(u_m^2)]$. To find the asymptotic form of the particular solution $\psi_p$, we first have,
\begin{equation}
\label{equ: asymptotic form for psi_r}
\psi_r(u_m) = 1- {2\over3}A_k u_m + O(u_m^2),
\end{equation}
\begin{equation}
\label{equ: asymptotic form for psi_s}
\psi_s(u_m) = {{1- 2A_ku_m + O(u_m^2)}\over{\sqrt{u_m}}}.
\end{equation}
Combining with above relations, it follows that $\psi_p$ becomes
\begin{equation}
\label{equ: asymptotic form for psi_p u_m <<1 and u_k << 1}
\psi_p(u_m) = \Psi_0 \phi_0 \Big [ -{2\over5}u_m^2 + {44\over525}A_ku_m^3 + O(u_m^4) \Big ].
\end{equation}
Finally, the denominator of Eq. (\ref{equ: comoving gauge psiDM density perturbation}) is
\begin{equation}
\label{equ: asymptotic form for denominator}
{1\over2}\Big [ \Big ({{d\Psi}\over{du_m}} \Big )^2 + \Psi^2 \Big ]= \Psi_0^2 \Big [ {1\over2} + O(u_m^2) \Big ],
\end{equation}
and the numerator
\begin{equation}
\label{equ: asymptotic form for numerator}
\left.\begin{aligned}
& {{d\Psi}\over{du_m}}{{d\psi_p}\over{du_m}} + \Psi\psi_p - \Big ({{d\Psi}\over{du_m}} \Big )^2\phi + {3\over2}{{{{d\Psi}\over{du_m}}\psi_p}\over{u_m}} \\
& = \Psi_0^2\phi_0 \Big [ \Big ( {8\over25} - {2\over5} - {4\over25} + {6\over25}\Big ){{u_m^2}} - {8\over175}A_k u_m^3 + O(u_m^4) \Big ]\\
& = \Psi_0^2\phi_0 \Big [ -{8\over175}A_ku_m^3 + O(u_m^4) \Big ],
\end{aligned}
\right.
\end{equation}
where the leading order terms cancel. We thus have
\begin{equation}
\label{equ: gauge covariant psiDM energy density asymptotic form}
\Delta_\psi = \phi_0\Big [ -{16\over175}A_ku_m^3 + O(u_m^4) \Big ] = \phi_0\Big [ -{12\over175}u_k^2 u_m^2 + O(u_m^4) \Big ].
\end{equation}
The gauge covariant $\psi$DM energy density perturbation $\propto u_k^2 u_m^2 \propto a^6$ when both $u_k$ and $u_m \ll 1$.

\bigskip

(B) Before mass oscillation and sub-horizon mode (phase (\rmnum{2}b))

\medskip

This phase is equivalent to $u_m \ll 1$ and $u_k \gg 1$, which imply $A_k \gg 1$. Unlike the previous case, $u_k$ is the natural variable here instead of $u_m$. The variables $\psi_r$, $\psi_s$ and the driving source $S$ will be expressed as functions of $u_k$.

From the definition of Kummer's function $M$, the homogeneous solutions $\psi_r$ and $\psi_s$ are
\begin{equation}
\label{equ: psi_r as function of u_k}
\psi_r = {{\sin(\sqrt{3}u_k)}\over{\sqrt{3}u_k}} + O(A_k^{-1}),
\end{equation}
\begin{equation}
\label{equ: psi_s as function of u_k}
\psi_s = 2\sqrt{A_k} \Big [{{\cos(\sqrt{3}u_k)}\over{\sqrt{3}u_k}} + O(A_k^{-1}) \Big ].
\end{equation}
On the other hand, the background field $\Psi$ has the asymptotic form $\Psi = \Psi_0[1-9A_k^{-2}u_k^{4}/80 + O(A_k^{-4})]$. Hence, the driving source $S$ becomes
\begin{equation}
\label{equ: source as function of u_k}
S = -\Psi_0\phi_0{6\over5} \Big [ 2{{\sin(u_k)}\over{u_k}} + {{\cos(u_k)}\over{u_k^2}} - {{\cos(u_k)}\over{u_k^3}} + O(A_k^{-2}) \Big ].
\end{equation}
Substituting Eqs. (\ref{equ: psi_r as function of u_k}), (\ref{equ: psi_s as function of u_k}) and (\ref{equ: source as function of u_k}) into Eq.(\ref{equ:psi_p}) and changing the variable from $u_m$ to $u_k$, the particular solution $\psi_p$ becomes
\begin{equation}
\label{equ: psi_p solution for u_m <<1 and u_k >> 1}
\psi_p = -{27\over10}\Psi_0\phi_0 A_k^{-2} [ u_k \sin(u_k) + O(1) +O(A_k^{-1})].
\end{equation}
Therefore, the gauge covariant energy density perturbation $\Delta_\psi$ is
\begin{equation}
\label{equ: Delta_psi asymtotic form u_m <<1 u_k >>1}
\Delta_\psi = \phi_0{81\over50} A_k^{-2} [ u_k^2 \cos(u_k) + O(u_k) +O(A_k^{-1}) ].
\end{equation}
The gauge covariant energy density perturbation grows as $u_k^2 \propto a^2$  modulating the oscillation of radiation perturbation.

\bigskip

(C) After mass oscillation and super-horizon mode (phase (\rmnum{1}a))

\medskip

In this phase, rather than dealing with Eq. (\ref{equ:psi_p}) directly we adopt an alternative approach to find the asymptotic solution.  This is because the Green's function integral of Eq. (\ref{equ:psi_p}) is quite cumbersome after mass oscillation. Before proceeding, the background fields  $\Psi_r$ and $\Psi_s$ in Eqs. (\ref{equ: Psi_r}) and (\ref{equ: Psi_s}) are redefined by dividing a factor $2^{1/4}\Gamma(5/4)$, i.e., $\Psi_r=J_{1/4}(u_m)/u_m^{1/4}$ and $\Psi_s = Y_{1/4}(u_m)/u_m^{1/4}$, to avoid irrelevant numerical factors appearing in the derivation. With this definition, the particular integral $\psi_p$ will also be divided by the same factor. This change does not affect the values of either $\delta_{\psi}$ or $\Delta_{\psi}$.

First, $\psi_p$ can be expanded as follows,
\begin{equation}
\label{equ:expansion of psi_p}
\psi_p(u_m, A_k) = \sum\limits_{n=0}^{\infty} \psi_p^{(n)}(u_m)A_k^n,
\end{equation}
for small $A_k$. Similarly, the same expansion applies to the driving source $S$,
\begin{equation}
\label{equ:expansion of source}
S(u_m, A_k) = \sum\limits_{n=0}^{\infty} S^{(n)}(u_m)A_k^n,
\end{equation}
The metric perturbation $\phi(u_k)$ inside $S(u_m, A_k)$ can also be expanded as
\begin{equation}
\label{equ: asymptotic form for phi}
\left.\begin{aligned}
\phi(u_k) & = 3\phi_0 \sum\limits_{n=0}^{\infty} {{(-1)^n}\over{(2n+1)!(2n+3)}}u_k^{2n} \\
          & = 3\phi_0\sum\limits_{n=0}^{\infty} {{(-{3\over4}u_m)^n}\over{(2n+1)!(2n+3)}} A_k^n.
\end{aligned}
\right.
\end{equation}
Substituting above relations into the perturbed field equation, it follows that,
\begin{equation}
\label{equ:equation of expanding method}
{{d^2\psi_p^{(n)}}\over{du_m^2}} + {3\over{2u_m}}{{d\psi_p^{(n)}}\over{du_m}} + \psi_p^{(n)} = \widetilde{S}^{(n)}, \text{  } n = 0,1,2,3,\cdots
\end{equation}
where $\widetilde{S}^{(n)} =  S^{(n)} - \psi_p^{(n-1)}/u_m$ with $\psi_p^{(-1)}(u_m)=0$ and $\psi_p^{(n)}(0)=d\psi_p^{(n)}/du_m(0)=0$ for any $n=0,1,2,3,\cdots$. It turns out that $\psi_p^{(n)}$ satisfies
\begin{equation}
\label{equ:solution of psi_p^n}
\psi_p^{(n)}(u_m) = \int_0^{u_m}{{\Psi_s(u_m)\Psi_r(x) - \Psi_r(u_m)\Psi_s(x)}\over{W(\Psi_r,\Psi_s)(x)}}\widetilde{S}^{(n)}(x)dx,
\end{equation}
with $W(\Psi_r,\Psi_s)(u_m) = 2u_m^{-3/2}/\pi$, which resembles Eq. (\ref{equ:psi_p}) since $\psi \rightarrow \Psi$ as $k \rightarrow 0$ or $A_k \rightarrow 0$.

Now, $\widetilde{S}^{(0)}(u_m) = -2\phi_0\Psi(u_m)$. Substituting it into Eq. (\ref{equ:solution of psi_p^n}), one obtains the two integrals in Eq. (\ref{equ:solution of psi_p^n}) for $n=0$,
\begin{equation}
\label{equ: solution Re hat psi 0}
\left.\begin{aligned}
& \int_0^{u_m} {{x^{3\over2}\Psi_s(x)\widetilde{S}^{(0)}(x)}\over{\Psi_0\phi_0}}dx = \\
& {{ 1-{}_1 F_2 \Big (-{1\over2};-{1\over4},{1\over4};-u_m^2 \Big ) }\over{4}} + \int_0^{u_m} {{x^{3\over2}\Psi_r(x)\widetilde{S}^{(0)}(x)}\over{\Psi_0\phi_0}}dx,
\end{aligned}
\right.
\end{equation}
and
\begin{equation}
\label{equ: solution Im hat psi 0}
\left.\begin{aligned}
& \int_0^{u_m} {{x^{3\over2}\Psi_r(x)\widetilde{S}^{(0)}(x)}\over{\Psi_0\phi_0}}dx = \\
& -{\pi\over4}u_m \Big [ 2u_mJ_{-{3\over4}}^2(u_m) - J_{-{3\over4}}(u_m)J_{{1\over4}}(u_m) + 2u_mJ_{{1\over4}}^2(u_m) \Big ],
\end{aligned}
\right.
\end{equation}
where ${}_1 F_2$ is the generalized hypergeometric function. Due to the fact $u_m \gg 1$ in this phase, we can have the following asymptotic forms,
\begin{equation}
\label{equ: asymtotic form for J_1/4}
 J_{{1\over4}}(u_m) =  \sqrt{{{2}\over{\pi u_m}}}\Big [ \cos\Theta +{3\over32}{{\sin\Theta}\over{u_m}} + O(u_m^{-2})\Big ], \text{ } u_m \gg 1,
\end{equation}
\begin{equation}
\label{equ: asymtotic form for Y_1/4}
Y_{{1\over4}}(u_m) = \sqrt{{{2}\over{\pi u_m}}} \Big [ \sin\Theta -{3\over32}{{\cos\Theta}\over{u_m}} + O(u_m^{-2})\Big ], \text{ } u_m \gg 1,
\end{equation}
\begin{equation}
\label{equ: asymtotic form for J_-3/4}
J_{-{3\over4}}(u_m) = \sqrt{{{2}\over{\pi u_m}}} \Big [ -\sin\Theta -{5\over32}{{\cos\Theta}\over{u_m}} + O(u_m^{-2})\Big ], \text{ } u_m \gg 1,
\end{equation}
\begin{equation}
\label{equ: asymtotic form for 1_F_2}
\left.\begin{aligned}
& {}_1 F_2 \Big (-{1\over2};-{1\over4},{1\over4};-u_m^2 \Big ) = \\
& -4u_m + 2\sqrt{2}\cos(2u_m) + O(u_m^{-1}), \text{ } u_m \gg 1,
\end{aligned}
\right.
\end{equation}
with $\Theta = u_m - 3\pi/8$. After straightforward substitution, $\psi_p^{(0)}$ becomes
\begin{equation}
\label{equ: asymtotic form for psi_p_0}
\sqrt{{\pi\over2}}{{u_m^{{3\over4}}\psi_p^{(0)}}\over{\Psi_0\phi_0}} = -\Big [ u_m \sin\Theta + {21\over32}\cos\Theta +O(u_m^{-1}) \Big ], \text{ } u_m \gg 1.
\end{equation}

It is difficult to get an analytic form for $\psi_p^{(n)}$ with $n =1,2,3,\cdots$. However, we can take advantage of $u_m \gg 1$, and separate the integral $\int_{0}^{u_m} = \int_{0}^{u_m^0} + \int_{u_m^0}^{u_m}$ where $u_m^0$ is a constant $\gg 1$. Decompose $\psi_p^{(n)}$ as follows:
\begin{equation}
\label{equ: separation of psi_p_n}
\left.\begin{aligned}
\psi_p^{(n)} = {\pi\over2} \Big [ & \Psi_s \int_{0}^{u_m^0} x^{{3\over2}}\Psi_r(x)\widetilde{S}^{(n)}(x)dx - \\
                                  & \Psi_r \int_{0}^{u_m^0} x^{{3\over2}}\Psi_s(x)\widetilde{S}^{(n)}(x)dx +\\
                                  & \Psi_s \int_{u_m^0}^{u_m} x^{{3\over2}}\Psi_r(x)\widetilde{S}^{(n)}(x)dx - \\
                                  & \Psi_r \int_{u_m^0}^{u_m} x^{{3\over2}}\Psi_s(x)\widetilde{S}^{(n)}(x)dx \Big ].
\end{aligned}
\right.
\end{equation}
The first two integrals give constants of order $O((u_m^0)^n)$, but the last two integrals are at least order of $O(u_m^n)\gg O((u_m^0)^n)$. Now, the last two integrations can be carried out via the above approximate formula (Eqs. (\ref{equ: asymptotic form for phi}), (\ref{equ: asymtotic form for J_1/4}), (\ref{equ: asymtotic form for Y_1/4}) and (\ref{equ: asymtotic form for psi_p_0})). The leading order forms of $\psi_p^{(n)}$ in the expansion of $u_m^{-1}$ can be obtained as follows:
\begin{equation}
\label{equ: solution of psi_p^n}
\left.\begin{aligned}
& -\sqrt{{\pi\over2}}{{u_m^{{3\over4}}}\over{\Psi_0\phi_0}}{{\psi_p^{(n)}A_k^n}\over3} = (-1)^n u_k^{2n}\Big \{ {2{u_m  \sin\Theta}\over{(2n+3)!}} - \\
& {3\over2} \Big [ {1\over{8(2n+3)!}} + {1\over{(2n+1)!(2n+3)}} + {1\over{2n(2n+1)!}}\Big ] \cos\Theta +  \\
& O \Big ( {1\over{u_m}} \Big ) \Big \}, \text{ } u_m \gg 1.
\end{aligned}
\right.
\end{equation}
Substituting Eqs. (\ref{equ: asymtotic form for psi_p_0}) and (\ref{equ: solution of psi_p^n}) into Eq. (\ref{equ:expansion of psi_p}) and summing up all order $A_k$, we arrive at
\begin{equation}
\label{equ: solution of psi_p}
\left.\begin{aligned}
& {{\psi_p}\over{\Psi_0}} = \phi_0\sqrt{{2\over\pi}}u_m^{-{3\over4}} \Big \{ 6\Big ( {{\sin(u_k)}\over{u_k^3}} -  {1\over{u_k^2}} \Big ) u_m \sin \Theta  - \\
& {9\over2} \Big [ {{\sin(u_k)}\over{u_k}} + {{8\cos(u_k)-1}\over{8u_k^2}} - {{7\sin(u_k)}\over{8u_k^3}} - Ci(u_k)   \\
& +\ln(u_k) + \gamma - {1\over2} \Big ] \cos \Theta + O\Big ({{\sum\limits_{l=0}^{\infty}a_l u_k^{2l}}\over{u_m}} \Big ) \Big \} \\
& =\phi_0 \Big \{ \Big [ 6\Big ( {{\sin(u_k)}\over{u_k^3}} -  {1\over{u_k^2}} \Big ) u_m + O \Big ( {{\sum\limits_{l=0}^{\infty}b_l u_k^{2l}}\over{u_m}} \Big ) \Big ]\Psi_s - \\
& {9\over 2} \Big [  {{\sin(u_k)}\over{u_k}} + {{\cos(u_k)}\over{u_k^2}} - {{\sin(u_k)}\over{u_k^3}} - Ci(u_k)\ + \ln(u_k) \\
& + \gamma - {1\over2} + O \Big ( {{\sum\limits_{l=0}^{\infty}c_l u_k^{2l}}\over{u_m}} \Big ) \Big ]\Psi_r \Big \} \\
& = \phi_0\Big [ 6\Big ( {{\sin(u_k)}\over{u_k^3}} -  {1\over{u_k^2}} \Big ) u_m + O \Big ( {{\sum\limits_{l=0}^{\infty}b_l u_k^{2l}}\over{u_m}} \Big ) \Big ]\Psi_s \\
& + {1\over2} \Big [ \Delta_C + \phi_0[9 \Big ( {{\sin(u_k)}\over{u_k^3}} -  {1\over{u_k^2}} \Big ) + O \Big ( {{\sum\limits_{l=0}^{\infty}c_l u_k^{2l}}\over{u_m}} \Big )] \Big ] \Psi_r \\
& \equiv \Im[\hat\psi]\Psi_s + \Re[\hat\psi]\Psi_r,
\end{aligned}
\right.
\end{equation}
where $a_l$, $b_l$ and $c_l$ are constants which make the infinite summation in Eq. (\ref{equ: solution of psi_p}) converges for $u_k \leq O(1)$. Here, we identify the coefficient of $\Psi_r$ to be $\Re[\hat\psi]$ and that of $\Psi_s$ to be $\Im[\hat\psi]$ discussed in phase (\rmnum{1}a) of the main text.

The energy density perturbation $\delta_\psi$ can be express as
\begin{equation}
\label{equ: asymtotic form for delta_psi u_m >>1}
\left.\begin{aligned}
\delta_\psi & = 2\Re[\hat\psi] - {{3\Im[\hat\psi]}\over{2u_m}} + {{3\Im[\hat\psi]}\over{u_m}}\sin^2 \Big ( u_m - {3\over8}\pi \Big ) \\
            & = \Delta_C + {{3\Im[\hat\psi]}\over{u_m}}\sin^2 \Big ( u_m - {3\over8}\pi \Big ) + O\Big ({{\sum\limits_{l=0}^{\infty}d_lu_k^{2l}}\over{u_m}} \Big )\phi_0.
\end{aligned}
\right.
\end{equation}
Similarly, the coefficient $d_l$ plays the same role with $a_l$, $b_l$ and $c_l$. From Eq. (\ref{equ: CDM gauge covariant energy density solution}), one can verify $\Delta_C \rightarrow 0$ as $u_k \rightarrow0$. This is consistent with the adiabatic condition discussed in the main text.

Finally, the asymptotic form of the gauge covariant energy density perturbation $\Delta_\psi$ is followed by adding the last term of the numerator in Eq. (\ref{equ: comoving gauge psiDM density perturbation}) to Eq. (\ref{equ: asymtotic form for delta_psi u_m >>1}). This additional term exactly cancel the second term in Eq.(\ref{equ: asymtotic form for delta_psi u_m >>1}) up to the leading order. Hence the gauge covariant energy density perturbation becomes
\begin{equation}
\label{equ: asymtotic form for Delta_psi u_m >>1}
\Delta_\psi  = \Delta_C + O\Big ({{\sum\limits_{l=0}^{\infty}e_lu_k^{2l}}\over{u_m}} \Big )\phi_0, \text{ } u_m \gg 1.
\end{equation}
Again, the constant $e_l$ makes the infinite summation in Eq. (\ref{equ: asymtotic form for Delta_psi u_m >>1}) converges when $u_k \leq O(1)$. It recovers the CDM result (Eq.(\ref{equ: CDM gauge covariant energy density asymptotic form})) to the leading term. The high order terms can be ignored as long as $A_k \ll 1$ or $u_k \leq O(1)$. Since $u_k \ll 1$ and $u_m \gg 1$ in this phase, this implies $A_k \ll 1$. Hence, the leading term of Eq. (\ref{equ: asymtotic form for Delta_psi u_m >>1}) is a good approximation, and the gauge covariant energy density perturbation of $\psi$DM follows the CDM case and grows as $u_k^2 \propto a^2$.

\bigskip

(D) After mass oscillation and sub-horizon mode (phase (\rmnum{1}b))

\medskip

The order of $A_k$ can be arbitrary in this phase and so high order terms in Eq. (\ref{equ: asymtotic form for Delta_psi u_m >>1}) cannot be ignored. We shall deal with Eq. (\ref{equ:psi_p}) directly. Noted that
\begin{equation}
\label{equ: boundness property of the integration}
\left.\begin{aligned}
& \lim\limits_{u_m\rightarrow\infty}\int_0^{u_m}{{\psi_r(x)}\over{W(\psi_r,\psi_s)(x)}}S(x)dx=C_1(A_k), \\
& \lim\limits_{u_m\rightarrow\infty}\int_0^{u_m}{{\psi_s(x)}\over{W(\psi_r,\psi_s)(x)}}S(x)dx=C_2(A_k),
\end{aligned}
\right.
\end{equation}
where $C_1$ and $C_2$ are constants depending only on $A_k$. This simplification is based on that the driving source $S(u_m)$ diminishes as $u_m\rightarrow\infty$, and the integrals in Eq.(\ref{equ: boundness property of the integration}) can be approximated by constants as long as $u_k \geq O(1)$ and $u_m \gg 1$.

From the asymptotic relations for $\psi_r$ and $\psi_s$,
\begin{equation}
\label{equ: asymtotic form of psi_r}
\left.\begin{aligned}
u_m^{{3\over4}}\psi_r \propto & {{e^{i \Big ( u_m - {3\over8}\pi + {A_k\over2}\ln(2u_m)\Big )}}\over{\Gamma \Big ( {3\over4} + {A_k\over2}i\Big )}} + {{e^{ -i \Big ( u_m - {3\over8}\pi + {A_k\over2}\ln(2u_m)\Big )}}\over{\Gamma \Big ( {3\over4} - {A_k\over2}i\Big )}}, \\
               & u_m \gg 1,
\end{aligned}
\right.
\end{equation}
\begin{equation}
\label{equ: asymtotic form of psi_s}
\left.\begin{aligned}
u_m^{{3\over4}}\psi_s \propto & {{e^{i \Big ( u_m - {1\over8}\pi + {A_k\over2}\ln(2u_m)\Big )}}\over{\Gamma \Big ( {1\over4} + {A_k\over2}i\Big )}} + {{e^{ -i \Big ( u_m - {1\over8}\pi + {A_k\over2}\ln(2u_m)\Big )}}\over{\Gamma \Big ( {1\over4} - {A_k\over2}i\Big )}}, \\
               & u_m \gg 1,
\end{aligned}
\right.
\end{equation}
one can obtain $\psi_p$ from Eq. (\ref{equ:psi_p}) in terms of the above expressions of $\psi_r$ and $\psi_s$ as well as $C_1(A_k)$ and $C_2(A_k)$. A straightforward calculation from Eq. (\ref{equ: comoving gauge psiDM density perturbation}) shows that the gauge covariant energy density perturbation can be expressed as
\begin{equation}
\label{equ: Delta_psi u_m >>1 u_k >>1}
\left.\begin{aligned}
\Delta_\psi \simeq & B_r(A_k) \cos (A_k \ln(u_k)) + B_s(A_k) \sin( A_k \ln(u_k) ), \\
                   & u_m \gg 1, \text{ } u_k \geq O(1),
\end{aligned}
\right.
\end{equation}
where the mass oscillation of $\psi_p$ is eliminated by that of $\Psi_r$ and it is left with only the matter-wave oscillation. Here $B_r(A_k)$ and $B_s(A_k)$ are the linear combination of $C_1(A_k)$ and $C_2(A_k)$. The gauge-covariant energy density oscillates with a constant frequency $A_k$ in $\ln(u_k)$ space.

Note that Equation (\ref{equ: Delta_psi u_m >>1 u_k >>1}) is valid for any $A_k$. One can identify constants $B_r$ and $B_s$ in the long-wave limit, i.e., $A_k \ll 1$, where Eq. (\ref{equ: Delta_psi u_m >>1 u_k >>1}) can be approximated as $\Delta_\psi \sim B_r(A_k)+B_s(A_k)A_k \ln(u_k)$. The coefficients $B_r(A_k)$ and $B_s(A_k)$ can be fixed by solution matching, which demands that Eq. (\ref{equ: Delta_psi u_m >>1 u_k >>1}) with $u_k \rightarrow 1$ from above should match Eq. (\ref{equ: asymtotic form for Delta_psi u_m >>1}) with $u_k \rightarrow 1$ from below. Comparing these two equations, one can find
\begin{equation}
\label{equ: B_r and B_s form for A_k <<1}
B_r(A_k) \rightarrow -9\phi_0\Big ( \gamma - {1\over2}\Big ),
B_s(A_k) \rightarrow -9\phi_0/A_k.
\end{equation}
This expression agrees with the result of Phase (\rmnum{1}b) in the main text.

\renewcommand{\theequation}{B\arabic{equation}}
\setcounter{equation}{0}
\section{Full treatment Evolution}
\label{app: full treatment evolution}

Governing equations are presented here with considerations of neutrino decoupling and photon-baryon coupling. For the zeroth order equations, the Friedmann equations with all relevant species are
\begin{equation}
\label{equ: Friedmann equations 1 with baryon and neutrino}
H^2 = {{8\pi G}\over{3}} a^2\Big ( \epsilon_{ph} +  \epsilon_\nu +  \epsilon_b + \epsilon_D \Big ),
\end{equation}
\begin{equation}
\label{equ: Friedmann equations 2 with baryon and neutrino}
\left.\begin{aligned}
H^{'}-H^2 = -4\pi G a^2 [ & ( \epsilon_{ph} + P_{ph} ) + ( \epsilon_\nu + P_\nu ) + \\
                          & ( \epsilon_b + P_b ) + ( \epsilon_D + P_D ) ].
\end{aligned}
\right.
\end{equation}
In addition, all species obey their own conservation laws (Eq. (\ref{equ: conservation})).

To enclose above equations, the equation of state is specified for the neutrino and baryon. Since neutrinos are relativistic particles, the equation of state is the same with photons, i.e., $P_\nu = \epsilon_\nu/3$.  Baryons are non-relativistic and described by a pressureless fluid, i.e., $P_b=0$.

First-order perturbed Einstein equations have the same structures as Friedmann equations, i.e.,
\begin{equation}
\label{equ: perturbed Einstein equation 1 wih baryon and neutrino}
-k^2\phi -3H(\phi^{'}+H\phi) = 4\pi Ga^2(\delta\epsilon_{ph}+\delta\epsilon_{\nu}+\delta\epsilon_b+\delta\epsilon_D),
\end{equation}
\begin{equation}
\label{equ: perturbed Einstein equation 2 wih baryon and neutrino}
\left.\begin{aligned}
\phi^{'}+H\phi = -4\pi Ga^2[& (\epsilon_{ph} + P_{ph})\theta_{ph} + (\epsilon_\nu + P_\nu)\theta_\nu +\\
                            & (\epsilon_b + P_b)\theta_b + (\epsilon_D + P_D)\theta_D].
\end{aligned}
\right.
\end{equation}

We need the continuity (density) equation and Euler (momentum) equation for every species to close perturbed equations. For $\psi$DM, it obeys Eqs. (\ref{equ: continuity equation}) and (\ref{equ: Euler equation}). On the other hand, neutrinos have decoupled from other species and become collisionless particles very early in the radiation-dominant era. Free-streaming damping will make its energy density perturbation diminished abruptly\cite{MaBertschinger1995} upon the mode entering horizon.  However, before entering horizon, neutrino perturbations have the same behavior as photon perturbations under the adiabatic condition. Hence the energy density and velocity potential perturbations of neutrinos can be approximated as,
\begin{equation}
\label{equ: delta_nu}
\delta_\nu =
\begin{cases}
\delta_{ph},  &\text{ } \tau \leq \tau_1, \\
0, &\text{ } \tau > \tau_1,
\end{cases}
\end{equation}
\begin{equation}
\label{equ: theta_nu}
\theta_\nu =
\begin{cases}
\theta_{ph},  &\text{ } \tau \leq \tau_1, \\
0, &\text{ } \tau > \tau_1,
\end{cases}
\end{equation}
where $\tau_1$ is the conformal time of the first oscillation null of $\Delta_{ph}(=\delta_{ph}-4H\theta_{ph})$ and is $k$-dependent\cite{MaBertschinger1995}.

The photon and baryon density and momentum equations read:
\begin{equation}
\label{equ: photon_continuity_equation_with_Thomson_scattering}
\delta_{ph}^{'} = {4\over3}k^2\theta_{ph} + 4\phi^{'},
\end{equation}
\begin{equation}
\label{equ: photon_Euler_equation_with_Thomson_scattering}
\theta_{ph}^{'} = -{1\over4}\delta_{ph} - \alpha Q -\phi,
\end{equation}
\begin{equation}
\label{equ: baryon_continuity_equation_with_Thomson_scattering}
\delta_b^{'} = k^2\theta_b + 3\phi^{'},
\end{equation}
\begin{equation}
\label{equ: baryon_Euler_equation_with_Thomson_scattering}
\theta_b^{'} + H\theta_b = \beta Q -\phi,
\end{equation}
where the subscripts "ph" and "b" stand for photon and baryon respectively, $Q\equiv \theta_{ph}-\theta_b$, $\beta = (4\epsilon_{ph}/3\epsilon_b)\alpha$ and $\alpha=l_T^{-1}$ with $l_T$ being the Thomson scattering mean free path. The last quantity is defined to be $l_T^{-1}\equiv a n_e \sigma_T$, where $n_e$ is the electron number density and $\sigma_T$ the Thomson cross-section. The photon and baryon fluids are coupled through the relative velocity potential $Q$ via Thomson scattering.

Note that $l_T^{-1}$ decreases with time as $a^{-2}$ for fully ionized baryons, and at some point $l_T^{-1}$ can be smaller than $k$.  After then, photons can no longer be described by a fluid and the above photon fluid equations fail to hold. Hence, for the analysis below to be valid we must demand $k l_T \ll 1$.    Moreover, we shall be focusing on modes that have already entered horizon,  $k/H \gg 1$. It turns out that $k_c$ of $m=10^{-22}$ eV satisfies the first criteria throughout the radiation dominant era and satisfies the second in the late epoch of the radiation era.

In the sub-horizon regime, the photon momentum equation is dominated by the photon pressure and the coupling term, and hence the gravity can be ignored. This situation is similar to the short wave limit of the self-gravitating nonrelativistic fluid where perturbations do not feel the gravity and become sound waves. We therefore drop $\phi$ in Eqs. (\ref{equ: photon_continuity_equation_with_Thomson_scattering}), (\ref{equ: photon_Euler_equation_with_Thomson_scattering}), (\ref{equ: baryon_continuity_equation_with_Thomson_scattering}) and (\ref{equ: baryon_Euler_equation_with_Thomson_scattering}).

Subtracting Eq. (\ref{equ: photon_Euler_equation_with_Thomson_scattering}) from Eq. (\ref{equ: baryon_Euler_equation_with_Thomson_scattering}), we can derive an equation for $Q$:
\begin{equation}
\label{equ: equation of Q}
Q^{'} = -{1\over4}\delta_{ph} + H\theta_{ph} - ( \alpha + \beta + H )Q.
\end{equation}
Substituting $(1/4)\delta_{ph}$ from Eq. (\ref{equ: equation of Q}) into Eq. (\ref{equ: photon_Euler_equation_with_Thomson_scattering}) and solving for $\theta_{ph}$ and $\theta_{ph}^{'}$, we arrive at
\begin{equation}
\label{equ: equation theta_ph 1}
\left.\begin{aligned}
\theta_{ph} = {3\over{k^2-3H{'}+3H^2}} \Big [ & H[Q^{'}+(\beta+H)Q]- \\
                                              & [Q^{'}+(\alpha+\beta+H)Q]^{'}\Big ],
\end{aligned}
\right.
\end{equation}
\begin{equation}
\label{equ: equation theta_ph 2}
\left.\begin{aligned}
\theta_{ph}^{'} = {{k^2-3H^{'}}\over{k^2-3H^{'}+3H^2}}\Big [ & Q^{'}+(\beta+H)Q + \\
                                                             & {{3H}\over{k^2-3H^{'}}}[Q^{'}+(\alpha+\beta+H)Q]^{'} \Big ].
\end{aligned}
\right.
\end{equation}

Now, since we are in the regime $k \gg H$, any term of order $O(H^2/k^2)$ or $O(H^{'}/k^2)$ can be ignored. However, we need to keep terms of order $O(H/k)$ as will become clear later. The reason for the coupling terms, such as $\alpha Q$ or $\beta Q$, to be finite leading-order terms is that we have large $\alpha$ and $\beta$ despite having a small relative velocity potential $Q$. Therefore, $Q^{'}$ is small compared to $\alpha Q$ or $\beta Q$, as $Q^{'} = O((k/\sqrt{3})Q)$ due to the photon oscillation to be shown below and furthermore $kl_T \ll 1$.

After the above considerations, we find
\begin{equation}
\label{equ: approximated equation theta_ph 1}
\theta_{ph} = - {3\over{k^2}}[ [Q^{'} + (\alpha+\beta)Q]^{'} - \{ H\beta Q \} ],
\end{equation}
\begin{equation}
\label{equ: approximated equation theta_ph 2}
\left.\begin{aligned}
\theta_{ph}^{'} = & {{\beta}\over{\alpha+\beta}}[Q^{'}+(\alpha+\beta)Q] + \\
                  & \Big \{ {{3H}\over{k^2}}[(\alpha+\beta) Q]^{'} + {{\alpha}\over{\alpha+\beta}}Q^{'} \Big \}.
\end{aligned}
\right.
\end{equation}
Terms in the brace brackets are terms of order $O(H/k)$.  If we ignore those high-order terms in Eqs. (\ref{equ: approximated equation theta_ph 1}) and (\ref{equ: approximated equation theta_ph 2}), it follows
\begin{equation}
\label{equ: equation of Q double prime}
[Q^{'}+(\alpha+\beta)Q]^{''} + {{k^2}\over3}{{\beta}\over{\alpha+\beta}}[Q^{'}+(\alpha+\beta)Q]=0.
\end{equation}

The quantity $Q^{'}+(\alpha+\beta)Q$ oscillates with the photon sound frequency modified by the presence of baryons, and this quantity is nothing more than $-(1/4)\Delta_{ph}$ from Eq. (\ref{equ: equation of Q}), where $\Delta_{ph}$ is the covariant energy density perturbation of photon. Thus the leading order terms give rise to a fast photon sound oscillation, and so the slow evolution of drag damping entails the next order correction to Eq. (\ref{equ: equation of Q double prime}), which all involve $Q^{'}$. To the leading order, $Q^{'}=ik(\beta/(3(\alpha+\beta)))^{1/2} Q$.

The approach by which we will derive the next order equation is similar to Phase (\rmnum{1}) of Sec. (\ref{sec: Passive Evolution and Asymptotic Solutions}), where we let $Q^{'}+(\alpha+\beta)Q = A(\tau) \exp[i k\int (\beta/(3(\alpha+\beta))) d\tau]$ and $A(\tau)$ is a slowly varying function of $\tau$. Substituting this expression into Eqs. (\ref{equ: approximated equation theta_ph 1}) and (\ref{equ: approximated equation theta_ph 2}), we find that
\begin{equation}
\label{equ: equation of amplitude}
{{dA}\over{d\eta}} =  -{{\alpha}\over{2(\alpha+\beta)}}\Big [ {{k^2}\over{3H(\alpha+\beta)}} + {1\over2} \Big ] A,
\end{equation}
where $\eta \equiv \ln a$.  This equation has an analytical solution.

It is clear from that Eq. (\ref{equ: equation of amplitude}) that for very long waves, $k^2\to 0$, the amplitude $A$ of $\Delta_{ph}$ will decrease when $\alpha/\beta (=(3/4)\epsilon_b/\epsilon_{ph})$ becomes non-negligible occurring shortly after radiation-matter equality. On the other hand for shorter waves where $k^2 > H(\alpha+\beta)$, the amplitude $A$ will further decrease and occur earlier in time.  We find that $k_c^2/H\beta \sim 1$ shortly before the radiation-matter equality and the drag damping begins to take effect for the critical mode near the end of radiation-dominant era.

\begin{figure}
\includegraphics[scale=0.35, angle=270]{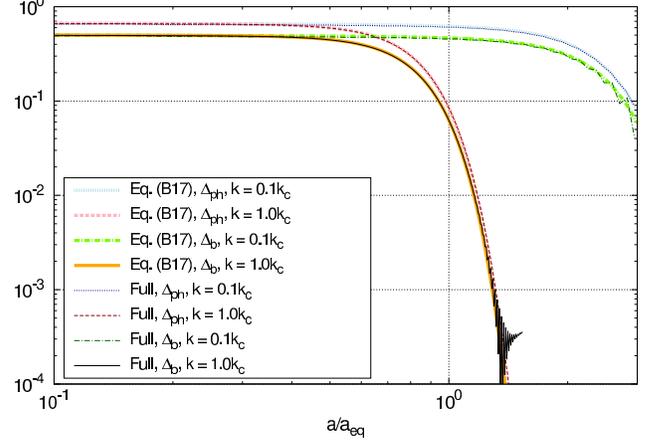}
\caption{The envelopes of gauge covariant photon and baryon energy density perturbations solved by Eq. (\ref{equ: equation of amplitude}) (thick line) and full treatment equations (thin line) for two distinct wavenumbers. Solutions of Eq. (\ref{equ: equation of amplitude}) well agrees with those of the full treatment.  Noted that damping for short wavelength modes is controlled by $(\epsilon_b/\epsilon_{ph})k^2l_T/H$, while damping for long wave modes is $k$-independent.}
\label{fig: compare_full_equation_Delta_photon}
\end{figure}

Figure (\ref{fig: compare_full_equation_Delta_photon}) shows the evolution of $\Delta_{ph}$ for $k=k_c$ and $k= 0.1 k_c$, where both solutions of Eq. (\ref{equ: equation of amplitude}) and of the full treatment are given. Good agreement is found. Baryon density perturbations can also be calculated from Eqs. (\ref{equ: baryon_continuity_equation_with_Thomson_scattering}) and (\ref{equ: approximated equation theta_ph 1}) with an understanding that $\Delta_b = \delta_b$ and $\theta_{ph} = \theta_b$ to the leading order.  It yields $\Delta_b = (3/4)\Delta_{ph}$. We also plot baryons solutions of the above approximate treatment and the full treatment. Again, good agreement is found, till after radiation-matter equality, as afterwards the gravity of matter can no longer be ignored.

\end{document}